\pdfoutput = 1

\documentclass[journal]{IEEEtran}
\ifCLASSINFOpdf
\else
\fi
\usepackage{bm}
\usepackage[cmintegrals]{newtxmath} % openbox clash with amsthm

\usepackage{epsfig}
\usepackage{graphicx}
\usepackage{subfigure}
\usepackage{placeins}
\usepackage{float}
\usepackage{boxedminipage}

\usepackage{color}

\usepackage{cite}
\usepackage{tabularx,colortbl}
\usepackage[ruled,vlined,linesnumbered]{algorithm2e}
\usepackage[cal=cm,bb=ams,scr=boondoxo] {mathalfa}

\usepackage{amsthm} % for theorem and assumption

\usepackage{etoolbox}% for space between author block and text body
\usepackage{mdwlist}
\usepackage{booktabs} % for topruls in table
\makeatletter
\patchcmd{\@maketitle} % for space between author block and text body
{\addvspace{0.5\baselineskip}\egroup} % upper space
{\addvspace{-1.5\baselineskip}\egroup} % lower space
{}
{}

\newtheorem*{assumption*}{\assumptionnumber}
\providecommand{\assumptionnumber}{}
\newtheorem*{definition*}{\definitionnumber}
\providecommand{\definitionnumber}{}


\begin{document}
	\title{\huge Parametric Sparse Bayesian Dictionary Learning for Multiple Sources Localization with Propagation Parameters Uncertainty and Nonuniform Noise}
	
	\author{
		Kangyong~You,~\IEEEmembership{Student~Member,~IEEE,}~Wenbin~Guo,~\IEEEmembership{Member,~IEEE},\\~Tao~Peng,~Yueliang~Liu,~Peiliang~Zuo,~and~Wenbo~Wang,~\IEEEmembership{Senior~Member,~IEEE}
		%and~Jane~Doe,~\IEEEmembership{Life~Fellow,~IEEE}% <-this % stops a space
		
			\thanks{Manuscript received Month Day, Year; revised Month Day, Year; accepted Month Day, Year. Date of publication Month Day, Year;
				date of current version Month Day, Year. This work was supported by the National Natural Science Foundation of China(61271181,61571054), the Science and Technology on Information Transmission and Dissemination in Communication Networks Laboratory Foundation.
				%and Research Fund for the Doctoral Program of Liaocheng University (Grant No. 318051835). 
				The associate editor coordinating the review of this
				manuscript and approving it for publication was Editor.  (\it Corresponding author: Wenbin Guo.)}
		%	
		%	\thanks{Kangyong You and Wenbin Guo are with the School of Information and Communication Engineering, Beijing University of Posts and Telecommunications, Beijing 100876, China, and also with the Science and Technology on Information Transmission and Dissemination in Communication Networks 	Laboratory, Shijiazhuang 050000, China (e-mail: ykyyiwang@bupt.edu.cn; 	gwb@bupt.edu.cn).}
		%	
		%%	\thanks{Lishan Yang is with Shandong Provincial Key Laboratory of Optical Communication Science and Technology, Liaocheng University, Liaocheng 252000, China(email: yanglishanbupt@163.com).}
		%	
		%	\thanks{ Tao Peng, Yueliang Liu, Peiliang Zuo and Wenbo Wang are  with the School of Information and Communication Engineering, Beijing University of Posts and Telecommunications, Beijing 100876, China (email:  pengtao@bupt.edu.cn; liuyueliang@bupt.edu.cn; zplzpl88@bupt.edu.cn;  wbwang@bupt.edu.cn);.}
		%%	\thanks{The authors are  with the School of Information and Communication Engineering, Beijing University of Posts and Telecommunications, Beijing 100876, China (email: ykyyiwang@bupt.edu.cn; 	gwb@bupt.edu.cn; zplzpl88@bupt.edu.cn; liuyueliang@bupt.edu.cn; wbwang@bupt.edu.cn);.}
		%	
		%	\thanks{Color versions of one or more of the figures in this paper are available online at http://ieeexplore.ieee.org.}

		\thanks{K. You and W. Guo are with the School of Information and Communication Engineering, Beijing University of Posts and Telecommunications, Beijing 100876, China, and also with the Science and Technology on Information Transmission and Dissemination in Communication Networks 	Laboratory, Shijiazhuang 050000, China (e-mail: \{ykyyiwang, gwb\}@bupt.edu.cn).}

		\thanks{ T. Peng, Y. Liu, P. Zuo and W. Wang are  with the School of Information and Communication Engineering, Beijing University of Posts and Telecommunications, Beijing 100876, China (email:  \{pengtao, liuyueliang, zplzpl88, wbwang\}@bupt.edu.cn.}
		
%		\thanks{This work was supported by the National Natural Science Foundation of China(61271181,61571054), the Science and Technology on Information Transmission and Dissemination in Communication Networks Laboratory Foundation.}
		
		%	\thanks{The authors are  with the School of Information and Communication Engineering, Beijing University of Posts and Telecommunications, Beijing 100876, China (email: ykyyiwang@bupt.edu.cn; 	gwb@bupt.edu.cn; zplzpl88@bupt.edu.cn; liuyueliang@bupt.edu.cn; wbwang@bupt.edu.cn);.}
		
		%\thanks{Color versions of one or more of the figures in this paper are available online at http://ieeexplore.ieee.org.}
		
	}

\maketitle
\vspace{-2cm}
\begin{abstract}
Received signal strength (RSS) based source localization method is popular due to its simplicity and low cost.
However, this method is highly dependent on the propagation model which is not easy to be captured in practice.
Moreover, most existing works only consider the single source and the identical measurement noise scenario, while in practice multiple co-channel sources may transmit simultaneously, and the measurement noise tends to be nonuniform.  
In this paper, we study the multiple co-channel sources localization (MSL) problem under unknown nonuniform noise, while jointly estimating the parametric propagation model. Specifically, we model the MSL problem as being parameterized by the unknown source locations and propagation parameters, and then reformulate it as a joint parametric sparsifying dictionary learning (PSDL) and sparse signal recovery (SSR) problem which is solved under the framework of sparse Bayesian learning with iterative parametric dictionary approximation. Furthermore, multiple snapshot measurements are utilized to improve the localization accuracy, and the Cram{\'e}r-Rao lower bound (CRLB) is derived to analyze the theoretical estimation error bound. Comparing with the state-of-the-art sparsity-based MSL algorithms as well as CRLB, extensive simulations show the importance of jointly inferring the propagation parameters, and highlight the effectiveness and superiority of the proposed method.
	
\end{abstract}

% to utilize information of multiple snapshot observations,
%  To analyze the accuracy of the proposed algorithms,


% \fi
\IEEEpeerreviewmaketitle

\section{Introduction}

\IEEEPARstart{L}{ocalization} has been attracting attention in many applications, across from commercial, industrial to defense areas, such as wireless networks, cognitive radio networks, spectrum monitoring, wireless sensor networks (WSNs), radar, and sonar \cite{niu_received-signal-strength-based_2018}. 
%It is also envisaged that most applications in the context of internet of things and 5G mobile networks depend on the location information to deliver better services. 
In particular, source localization in WSNs has far-reaching applications\cite{Win_Network-loc-navi-via-cooperation_2011,Lymbe_Micro-indoor-loc-compt_2017}, where WSNs consist of a large number of cheap, densely deployed sensors with limited sensing and communication abilities, which monitor a spatial physical phenomenon (e.g. temperature, sound intensity, radio signal intensity, pollution concentrations, etc.) and regularly report their measurements to a Fusion Center (FC).
%\cite{Nevat_Random-field-reconstr-with-Quanti-WSN_2013,Nevat_Distr-Detec-WSNs-FC_2014}. 

According to the information available for localization in time domain, frequency domain, angular domain, and energy domain, several representative source localization methods have been proposed over the past years, such as time of arrival (TOA)\cite{Cheung_LS_TOA_2004}, time difference of arrival (TDOA)\cite{Wang_TDOA_2017},  frequency difference of arrival (FDOA)\cite{TDOA_FDOA2}, direction of arrival (DOA) \cite{Godara_Application_AA_1997,Stein_CaSCADE_2017} and RSS based localization algorithms \cite{zuo1, zuo2 ,Shenxiaohong,Bandiera_A-cogitive-alg-RSS-loc_2015,Yan_A-framwork-low-complexity-LS-loc_2010}. In these methods, sophisticated ones are often with high accuracy but pay the price of advanced radio receiver, processing, and communication abilities, e.g., DOA approach for narrowband signal sources requires multiple antennas or antenna array, while TOA, TDOA, or FDOA for wideband signal sources face the challenges of timing synchronization, coherent demodulation, and  high-speed analog-to-digital conversion (ADC) (especially when ultra-wideband signal are interested \cite{UWB}). Besides, DOA, TOA, TDOA, and FDOA are very sensitive to the availability of line of sight (LOS).
On the contrary, RSS measurements, operating in both LOS and non-LOS (NLOS) environments and readily available from any radio interface, are simple and require no additional sensor functionalities. As a result, RSS-based source localization approaches have gained popularity in WSNs where simplicity, low energy consumption and low cost are the main requirements.

\vspace{-1em}
\subsection{Related Works}

In the past decades, many RSS-based source localization approaches have been proposed (see the overviews in \cite{Mao_Suvy-WSN-Loc_2007,Nguyen_A_Bayes-MultiSource-loc-WSN_2016,niu_received-signal-strength-based_2018}).
%Given the sensors' locations and the propagation model between sources and sensors, the localization problem can be cast as an optimization problem with constraints on the RSS observations of the sensors. 
Early literature devotes to single source localization (SSL). In the early ages of RSS-based SSL, range-based localization was achieved through trilateration \cite{trilateration} or multilateration algorithms  \cite{ multilateration}. These techniques are simple but suboptimal, and their accuracy is also limited. The maximum likelihood estimate (MLE) based approaches \cite{ML-loc-single-source-2003,ML-loc-single-source-2010} are more accurate but highly nonlinear,  nonconvex and exhausted to search for the global maximum.  Recently,  there has been an increasing interest in relaxing the MLE problem, such as  algorithms based on the linear least squares(LLS) \cite{DLi_Energy-based-collaborative-loc_2003,LLS-single-loc-2011}, the projection onto convex sets \cite{gholami2011wireless-POCS} and the semidefinite programming (SDP) \cite{Wang_a-new-app-node-loc_2011}. 

Later, more and more efforts are focusing on MSL where energy information of multiple co-channel sources  are coupled in RSS measurement since they share the same time and frequency resources. This phenomenon exists extensively in many applications, such as acoustic sources localization where multiple sources may make sounds simultaneously, spectrum monitoring  where an illegal radio occupies the legal user's frequency band, cognitive radio  where primary users and secondary users share the same time and frequency resources.  Moreover, with the rapid advancement of 5G communication, non-orthogonal multiple access (NOMA) techniques  and 5G enabled Internet-of-things (IoT) applications \cite{IoT} will make this phenomenon more ubiquitous. In the multiple co-channel sources scenario, localization problem turns tougher and more challenging, while the  aforementioned SSL methods fail to make it.

To locate multiple sources, the region of interest (ROI) is usually discretized into a set of grid points (GPs) as searching space (or location candidates). MLE approach was first proposed in \cite{Shenxiaohong} where a combination of multiresolution search algorithm and expectation-maximum (EM)-like algorithm was used to perform exhausted coordinate search along each dimension in searching space. Later, to reduce the computation cost, and to improve the estimation performance as well as robustness in the presence of noise and small observation size, spatial sparsity based approaches have been gradually gaining popularity \cite{Feng_2009,Cevher_2008,Feng_2010,Zhang_2011,Feng_2012}. The main idea is that assume sources are located on the predefined GPs, and then under specific conditions \cite{CS}, multiple source locations can be estimated by searching the sparsest solution of an underdetermined linear localization equation \cite{Cevher_2008}. Nevertheless, sources may deviate from the predefined GPs  (off-grid) in reality, which will impair the localization performance greatly. In compressive sensing (CS) theory  \cite{CS}, off-grid sources bring basis mismatch problem which can not be eliminated by finer grid  granularity \cite{BM}. Recently, some methods have been proposed to address the off-grid sources localization problem \cite{SUN_TC_2017,SUN_CL_2017,GEMTL}.

However, a major challenge for RSS-based localization lies in the uncertainty of propagation model. All of aforementioned works assume the characteristics of the  propagation model are known and given. Nevertheless,
the propagation model in practical application is not easy to be captured with time-varying propagation environment.
Generally, the propagation process is characterized by some propagation parameters, such as the path-loss exponent (PLE) and transmitted powers.  
The single source localization problem with unknown propagation parameters has been addressed in {\cite{PLESingle2,PLESingle1,PLESingle3,Gholami2}}. 
In \cite{PLESingle2}, a linear regression model was proposed for PLE estimate, and the total least squares (TLS) method was exploited to infer the unknown PLE.
In {\cite{PLESingle1}, a Bayesian minimum mean square error (MMSE) estimator was developed to locate the source with unknown PLE. In \cite{PLESingle3}, semidefinite programming (SDP) relaxation technique was adopted to estimate the transmitted power of the source. In \cite{Gholami2}},  the source was located with unknown PLE and unknown transmitted  powers through solving  a general trust region problem. Nevertheless, the MSL problem has not been well addressed with unknown propagation parameters. 

\vspace{-1em}
\subsection{Contributions}

In this paper, we extend the RSS-based SSL work of  \cite{Gholami2} to locate multiple co-channel sources in the presence of uncertain path-loss exponent and unknown transmitted powers. Moreover, we consider the more general case of nonuniform measurement noise and multiple snapshots model. To this end, an efficient  parametric sparse Bayesian dictionary learning (PSBDL) algorithm is proposed. The main contributions of this paper are summarized as follows.

\begin{enumerate}
	\item To the best of our knowledge, we first provide a unified framework to locate multiple sources while jointly inferring the propagation parameters utilizing spatial sparsity. Specifically, we provide a localization model parameterized by source locations and propagation parameters. Then, we propose an approximation model to learn the  sparsifying parameterized localization dictionary.
	\item Under the proposed localization model, we reformulate the MSL problem as a joint PSDL and SSR problem which is effectively solved by incorporating the proposed parametric dictionary approximation model with multiple measurement vector (MMV) sparse Bayesian learning framework.
	\item We provide CRLB analysis for the considered problem, and compare the proposed method with the state-of-the-art spatial sparsity based MSL methods. Extensive simulations show the importance of jointly estimating the propagation parameters, and highlight the effectiveness of the proposed framework.
	%	 Meanwhile, some important guidelines for future research are discussed.  
\end{enumerate}

The remainder of this paper is organized as follows: Section II first  presents the proposed localization dictionary model and  parameterized  dictionary approximation model,  and  then  reformulate the MSL problem. Section III is devoted to developing the proposed PSBDL algorithm. Section IV elaborates on the derivation of the  CRLB.  Numerical simulation results are reported in Section V. Discussion is presented in Section VI. Section VII closes this paper with conclusions.

{\it Notation:}  $x_i$ is the $i$-th entry of a vector $\bm{x}$. $\bm{A}_i$, $\bm{A}^i$, and ${A}_{i,j}$ are the $i$-th column, $i$-th row, and $(i,j)$-th entry of  a matrix $\bm{A}$. $\lVert \cdot \rVert_0$, $\lVert \cdot \rVert_1$, $\lVert \cdot \rVert_2$, and $\lVert \cdot \rVert_F$   denote the pseudo-$\ell_0$ norm, $\ell_1$ norm, $\ell_2$ norm, and Frobenius norm, respectively. $(\cdot)^{T}$ denotes transpose operator. $\operatorname{tr}(\cdot)$ and $\lvert \cdot \rvert$ denote the trace and determinant operator, respectively. $\operatorname{diag}(\bm{x})$ is a diagonal matrix with vector $\bm{x}$ being its diagonal elements. $\operatorname{diag}(\bm{A})$ denotes a column vector composed with the diagonal elements of matrix $\bm{A}$. $\circ$ is the Hadamard (element-wise) product operator. For clear and concise presentation, some functions are abbreviated sometimes by omitting the input variables in context, e.g.  $\bm{\Phi}(\bm{\theta})$ is abbreviated as $\bm{\Phi}$, and ${f(\bm{s}_i,\bm{t}_k,\gamma)}$ abbreviated as $f$. $\bm{1}_N$ and $\bm{I}_N$ denote the all ones vector and  the identity matrix of dimension $N$, respectively.

\section{Problem Formulation}
%\vspace{-0.4cm}
In this section, we first revisit the fundamentals of sparsity-based MSL problem, and present the proposed localization model considering both the 
unknown source locations and the unknown propagation parameters. Then, we proposed a parameterized dictionary approximation model and  reformulate the MSL problem as a joint PSDL and SSR problem.

\vspace{-0.2cm}
\subsection{MSL Model}  \label{Section:MSL_model}
%\vspace{-1em}
The system of consideration consists of $K$ sources with unknown locations { \small $\!\!\!\mathcal{T} \!\!=\!\!\left\{\bm{t}_{k}\!=\!\left[u_{k}^{t}, v_{k}^{t}\right], k\!=\!1, \!\cdots\!, K\right\}\!\!$} and $\!M\!$ passive sensors with known locations {\small $\!\mathcal{S}\!\!=\!\!\left\{\bm{s}_{i}\!=\!\left[u_{i}^{s}, v_{i}^{s}\right], i\!=\!1, \!\cdots\!, M\right\}$} in a two-dimensional ROI with $u$ and $v$ being the Cartesian coordinates. The RSS measurement of the $i$-th sensor at time snapshot $t$ can be expressed as \cite{Shenxiaohong}
\begin{equation}\label{SignalModel:yi_t}
	{y_i}(t) = \sum \nolimits_{ k = 1}^K {P_k}(t) {f(\bm{s}_i,\bm{t}_k,\gamma )} + \varepsilon_i(t),
\end{equation}
where $\varepsilon_i(t)$, $f(\cdot)$,  $P_k(t)$, and $\gamma$  are the unknown measurement noise of sensor $i$ at time $t$, the propagation model, the transmitted power of source $k$ at a reference distance $d_0$  at time $t$, and the PLE, respectively. Generally, the PLE varies from 2 (free space) to 6 (e.g., some indoor scenario) \cite{Gholami2}, and is off-line calibrated in conventional routine. The  matrix-vector formulation of the single measurement vector (SMV) signal model for time $t$ is:
\vspace{-0.1cm}
\begin{equation}\label{SignalModel:y_t}
	\bm{y}(t) =  \bm{\Phi}(\mathcal{T},\gamma){\bm{\omega}}(t)   + \bm{\epsilon}(t),
%	\vspace{-0.2cm}
\end{equation}
with { $\bm{\epsilon}(t) \!=\! \left[ \varepsilon_1(t),\!\cdots\!,\varepsilon_M(t)\right]^{ T} $, $\bm{y}(t) \!=\! \left[ y_1(t),\!\cdots\!,y_M(t)\right]^{ T} $, $\bm{\omega}(t) \!=\! \left[ P_1(t),\!\cdots\!,P_K(t)\right]^{ T}$,  $\bm{\Phi}( \mathcal{T},\gamma )_{i,k} \!=\! {f(\bm{s}_i,\bm{t}_k,\gamma )} $}.

We further consider there are $T$ snapshots RSS measurement available, denote $\bm{Y} \!=\! \left[\bm{y}(1),\cdots,\bm{y}(T)\right]$, $\bm{W} \!=\! \left[{\bm{\omega}}(1),\cdots,{\bm{\omega}}(T)\right]$ and $\bm{E} \!=\! \left[\bm{\varepsilon}(1),\cdots,\bm{\varepsilon}(T)\right]$, and then the SMV  model in (\ref{SignalModel:y_t}) evolves into the MMV model as
\vspace{-0.1cm}
\begin{equation}\label{SignalModel:Y}
	{\bm{Y}} = \bm{\Phi}(\mathcal{T}, \gamma) \bm{W} + \bm{E },
\end{equation} 
with $\bm{Y},\bm{E} \!\in\! \mathbb{R}^{M \times T}$, $\bm{W} \!\in\! \mathbb{R}^{K \times T}$, and $\bm{\Phi}(\mathcal{T}, \gamma) \!\in\! \mathbb{R}^{M \times K}$. Thus, the SMV signal model in (\ref{SignalModel:yi_t}) is a special case when $T=1$.

Generally, the noise statistics of the sensors observations are different. Thus, we assume $\bm{\varepsilon}(t)$ is a nonuniform noise.
As a result, the MSL task can be summarized as given the measurement matrix $\bm{Y}$, sensor location set $\mathcal{S}$, and parametric propagation model $ {f(\bm{s}_i,\bm{t}_k,\gamma)}$, how to  infer the source location set $\mathcal{T}$ in the presence of unknown nonuniform noise $\bm{E}$ and unknown propagation parameters $\gamma$ and $P_k \text{, for } k=1,\dots,K$.

%\vspace{-1em}
\subsection{Traditional Spatial Sparsity Based MSL Methods}

To alleviate the problem difficulty, traditional sparsity-based methods ( \cite{Cevher_2008, Feng_2009, Feng_2010, Feng_2012, Zhang_2011}, etc.) assume that the PLE is precisely known, and all sources are located on predetermined candidate GP set { \small $\mathcal{G} \!=\!\left\{\bm{g}_{j} \!=\!\left[u_{j}, v_{j}\right], j\!=\!1, \!\cdots,\! N\right\}$},i.e. {\small ${\mathcal T} \!\subset\! {\mathcal G}$}.  

Assume the sources are static during the observation period, then $\bm{Y}$ has sparse representation in a localization dictionary $\!\bm{\Phi}(\cal G)\!$ with the fact that $ K \ll N $.  Therefore, the MSL model in (\ref{SignalModel:Y})  can be cast into a standard  sparse recovery model as
\begin{equation}\label{CSModel}
	\bm{Y} = \bm{\Phi}({\mathcal G}) \bm{X} + \bm{E},
\end{equation}
where  $\bm{X} = [\bm{x}(1),\cdots,\bm{x}(T)]$  and  $\bm{X}_{i,t} = P_k(t)$ when source $k$ locates on GP $i$, and otherwise $\bm{X}_{i,t} = 0$. Thus $\bm{X}$ is a common sparse (or row-sparse) coefficient matrix \cite{ComomSparse}, i.e., all the columns $\bm{X}_t$ share the same sparse support. As a result, the row support of $\bm{X}$ encodes the source locations in candidate GP set $\mathcal{G}$ and the corresponding rows in $\bm{X}$ encode the transmitted powers in different time snapshots.

In this way, localization can be  transformed into a standard MMV row-sparse recovery problem as
\begin{equation}\label{CS_MSL}
	\hat{\bm{X}} \!=\! \mathop {\arg\min }\limits_{\bm{X}}{\mathcal{R} \left( X\right)}, \;\text{s.t.}\;{\lVert \bm{Y} \!-\! \bm{\Phi}(\mathcal{G}) \bm{X}  \rVert}_F < \epsilon,
	%\vspace{-0.05cm}
\end{equation}
where $\epsilon$ bounds the amount of noise in $\bm{Y}$,  and  $\mathcal{R} \left( X\right)$ denotes the row sparsity of $\bm{X}$, i.e., the number of non-zero rows. Problem (\ref{CS_MSL})  can be solved using standard MMV compressive sensing methods, such as S-OMP\cite{SOMP}, M-BP\cite{MBP}, M-FOCCUS\cite{MFOCUSS}, M-SBL\cite{MSBL}, etc. In particular, details about traditional sparsity-based MSL when $T=1$ are referred to  \cite{Cevher_2008, Feng_2009, Feng_2010, Feng_2012, Zhang_2011}.

\vspace{-0.2cm}
\subsection{The Proposed Parametric Dictionary Model and Its Approximation}\label{Section:II:ApproxModel}

In practice, source locations may deviate from the predefined candidate GPs and the off-line calibrated path-loss exponent may differ from that in on-line RSS measurement. Thus, it is more realistic and important to treat  the candidate GP set  $\mathcal{G}$ and the PLE $\gamma$ as unknown variables to be inferred from the on-line RSS measurements. To this end, the localization dictionary is modeled as $\bm{\Phi}(\mathcal{G},\gamma)$. Accordingly, the MSL model (\ref{CSModel}) evolves into 
\begin{equation}\label{AGModel}
	\setlength\abovedisplayskip{1pt}
	\setlength\belowdisplayskip{1pt}
	\bm{Y} = \bm{\Phi}(\mathcal{G}, \gamma) \bm{X} + \bm{E},
\end{equation}
and the corresponding optimization problem turns into 
\noindent the following joint PSDL and SSR problem
\begin{subequations} \label{JointOpt}
	\setlength\abovedisplayskip{1pt}
	\setlength\belowdisplayskip{1pt}
	\begin{align}
		&\left( \hat{\bm{X}}, \hat{\bm\Phi} \right) = \mathop {\arg\min }\limits_{\bm{X},\bm{\Phi}}{\mathcal{R} \left( X\right)} \\
		&\text{s.t.}\quad{\lVert \bm{Y} - \bm{\Phi}(\mathcal{G},\gamma) \bm{X}  \rVert}_F < \epsilon,
	\end{align}
\end{subequations}

However, to infer   $\bm{\Phi}(\mathcal{G},\gamma)$ directly is nearly impossible since the goal function w.r.t. the  dictionary parameters $\mathcal{G}$ and $\gamma$ is highly nonconvex. As a result, some approximation methods must be resorted to. Have in mind that in the implementation of an iterative algorithm, the dictionary parameters are often initialized with $\mathcal{G}^{(0)}$ and $\gamma^{(0)}$ to construct an  initial dictionary which will be updated in the subsequent inference. 
Thus, denote by $\bar{\mathcal{G}}$ the proper candidate GP set satisfying $\mathcal{T} \subset \bar{\mathcal{G}}$, $\bar{\gamma}$ the true PLE, $\bm{\delta}_g = [\bm{\delta}_{u}, \bm{\delta}_{v}]$ the grid offset to  $\bar{\mathcal{G}}$ of the current grid  estimation $\mathcal{G}^{(k)}$,  $\delta_{\gamma}$ the PLE offset to $\bar{\gamma}$ of the current PLE estimation ${\gamma}^{(k)}$, we can expand the dictionary by each entry using Taylor series, and approximate it through keeping the linear parts as
\vspace{-0.1cm}
\begin{align}
	\setlength\abovedisplayskip{1pt}
	\setlength\belowdisplayskip{1pt}
	\!\!\!\!&\bm{\Phi}\left(\bar{\mathcal{G}},\bar{\gamma} \right) \!\approx  \bm{\Phi}_0  +  \bm{\Phi}^{\prime}_{{u}} \!\!\left(\mathcal{G}^{(k)},\gamma^{(k)}  \right) \operatorname{diag}\left( \bm{\delta}_u\right) \notag\\
	&\;+  \bm{\Phi}^{\prime}_{{v}} \!\!\left(\mathcal{G}^{(k)},\gamma^{(k)}  \right)\operatorname{diag}\left( \bm{\delta}_v\right) +  \delta_{\gamma}\bm{\Phi}^{\prime}_{\gamma} \!\!\left(\mathcal{G}^{(k)},\gamma^{(k)}  \right) 
	\vspace{-0.5cm}	 
\end{align}
where $\bm{\Phi}_0 \!=\! \bm{\Phi}\left( \mathcal{G}^{(k)},\gamma^{(k)} \right)$  and for $\chi = u,v,\gamma$, $\bm{\Phi}^{\prime}_{\chi}$ is the partial differential matrix with the $(i,j)$-th entry being the partial differential item expressed as $ \left(\bm{\Phi}^{\prime}_{\chi}\right)_{i,j} = \partial {f(\bm{s}_i,\bm{g}_j,\gamma )} / \partial \chi $. 

\vspace{-0.2cm}
\subsection{Problem Reformulation}

Based on above approximation model, we can relax and solve the joint optimization problem (\ref{JointOpt})  iteratively. In each iteration, given current dictionary parameter as $\mathcal{G}^{(k)}$, $\gamma^{(k)}$, we have to settle  the following joint PSDL and SSR subproblem 
\begin{subequations} \label{iter_subq}
	\begin{align}
		&\left( \hat{\bm{X}}, \hat{\bm{\delta}}_g,\hat{\delta}_{\gamma} \right) = \mathop {\arg\min }\limits_{\bm{X},\bm{\delta}_g, \delta_{\gamma}}{\mathcal{R} \left( X\right)} \label{iter_subq:goal}\\
		&\text{s.t.} \notag\\
		& \bm{\Phi}_0 = \bm{\Phi}\left( \mathcal{G}^{(k)}, \gamma^{(k)} \right), \label{iter_subq:c1}\\
		&\bm{\Phi} \!=\! \bm{\Phi}_0  \!+\!  \bm{\Phi}^{\prime}_{{u}} \!\!\left( \mathcal{G}^{(k)}, \!\gamma^{(k)}  \right)\! \operatorname{diag}\left( \bm{\delta}_u\right)\!+\!  \delta_{\gamma}\bm{\Phi}^{\prime}_{\gamma} \!\!\left( \mathcal{G}^{(k)}, \gamma^{(k)}  \right) \notag \\ 
		&\quad\;+\!  \bm{\Phi}^{\prime}_{{v}} \!\!\left( \mathcal{G}^{(k)},\! \gamma^{(k)}  \right) \!\operatorname{diag}\left( \bm{\delta}_v\right), \label{iter_subq:c2}\\
		&{\lVert \bm{Y} - \bm{\Phi}\bm{X}  \rVert}_F < \epsilon,\label{iter_subq:c3}\\
		& \bm{\delta_{u} \!\in\! [\bm{LB_u}, \bm{UB_u}]}, \bm{\delta_{v}} \!\in\! [\bm{LB_v},\bm{UB_v}], \delta_{\gamma} \!\in\! [LB_{\gamma}, UB_{\gamma}]
	\end{align}
\end{subequations}
with $LB_{\chi}$,$UB_{\chi}$  being the lower and upper boundary for $\bm{\delta}_{\chi}, \chi = u, v,\gamma$, respectively.

Once problem (\ref{iter_subq}) is solved, we can update the dictionary parameters simply as 
\begin{equation}
	\mathcal{G}^{(k+1)} =  	\mathcal{G}^{(k)} + \hat{\bm{\delta}}_g, \quad \gamma^{(k+1)} = \gamma^{(k)} + \hat{\delta}_{\gamma}, 
\end{equation}
and then solve the subproblem again until it converges.

\section{Parametric Sparse Bayesian  Dictionary Learning for Multiple Sources Localization }
%\vspace{-0.2cm}
In this section, we are devoted to solving  problem (\ref{iter_subq}) from the perspective of probabilistic inference.
First, a  hierarchical sparsity-promoting probabilistic model is imposed for model (\ref{AGModel}). Then, problem (\ref{iter_subq}) is solved based on Bayesian inference. At last, the proposed PSBDL algorithm is summarized, and its complexity is discussed.

\vspace{-0.1cm}
\subsection{ Hierarchical Sparse Probabilistic Model} \label{Section:III:BayesModel}
The hierarchical probabilistic model is expressed as
\begin{subequations}\label{HPM}
	\begin{align}
		\setlength\abovedisplayskip{1pt}
		\setlength\belowdisplayskip{1pt}
		&\bm{E}\mid \bm{\beta}  \sim \prod\nolimits_{t=1}^{T}{\mathcal{N}}\left({\bm{\varepsilon}(t)} \mid 0, \operatorname{diag}(\bm{\beta})^{-1} \right), \label{NoiseModel1}
		\\
		&\bm{\beta};a,b \sim \prod\nolimits_{j=1}^M Gamma \left( \beta_j \mid a,b \right),\label{NoiseModel2}
		\\
		& \bm{X} \mid \bm{\alpha} \sim \prod\nolimits_{t=1}^{T}{\mathcal N}\left({\bm{x}(t)}|0, \operatorname{diag}(\bm{\alpha}) \right), \label{Xstage1} \\
		&  \bm{\alpha};\lambda \sim \prod\nolimits_{i = 1}^N {Gamma \left({\alpha_i} \mid 1,\frac{\lambda}{2} \right)}, \label{Xstage2}\\
		& \gamma \sim {\it{Uniform}}\left(\gamma\mid 2,6\right),
	\end{align}
\end{subequations}
where the probability density function (PDF)  of a multivariate Gaussian distribution random variable $\bm{x}$ with mean $\bm{\mu}$ and covariance $\bm{\Sigma}$ is 
\begin{equation}\label{GaussPDF}
	\setlength\abovedisplayskip{1pt}
	\setlength\belowdisplayskip{1pt}
	{\mathcal{N}\!\left(\bm{x}|\bm{\mu},\bm{\Sigma}\right)} \!=\! \frac{1}{\sqrt{(2\pi)^{N} \lvert\bm{\Sigma}}\rvert} \exp \! \left\{\!-\frac{(\bm{x} \!-\! \bm{\mu})^{ T}{\bm \Sigma^{-1}
		}(\bm{x} \!-\! \bm{\mu})}{2}\right\},
	\vspace{0.1cm}
\end{equation}
the PDF of a Gamma distribution random variable $x$ with shape parameter $a$ and rate parameter $b$ is
\begin{equation}
	Gamma (x ; a,b) =\Gamma(a)^{-1}b^{a} x^{a-1} \exp\left\{ -b x\right\}
\end{equation}
with $\Gamma(\cdot)$ being the Gamma function, the PDF of a uniform distribution random variable $x$ in the interval of $[a,b]$ is
\begin{equation}
	\setlength\abovedisplayskip{0.1pt}
	\setlength\belowdisplayskip{0.1pt}
	{\it{Uniform}}(x ; a,b) = \frac{1}{b-a}.
\end{equation} 

\vspace{0.2cm}
Intuitively, for $t=1,\cdots, T$, noise $\bm{\varepsilon}(t)$ is independent identically distributed (i.i.d.) nonuniform Gaussian random variables whose variance is governed by the conjugate hyperprior shown in (\ref{NoiseModel2}). Moreover, all columns of $\bm{X}$ are independent and share the same prior which is shown in \cite{BCS_Laplace}  to be a Laplace distribution as
\vspace{-0.2cm}
\begin{align}\label{LaplacePDF}
	\setlength\abovedisplayskip{0.5pt}
	p\left(\bm{x}(t);\lambda\right) &=  \int p\left(\bm{x}(t) | \bm{\alpha} \right) 	p\left(\bm{\alpha} ; \lambda \right)\,  d \bm{\alpha} 
	\notag\\
	&= \frac{\sqrt{\lambda}}{2}\exp\left\{-\sqrt{\lambda}{\lVert\bm{x}(t)\rVert_1}\right\}.
\end{align}

Above Laplace distribution is also termed as Bayesian LASSO \cite{tibshirani1996} whose counterpart in optimization theory, LASSO, is the best convex approximation to the $\ell_0$-norm. The distribution in (\ref{LaplacePDF}) is strongly peaked at the origin, thus it is a sparse prior that favors most entries of vector $\bm{x}(t)$ being zeros. Since all columns of $\bm{X}$ are governed by the same sparse prior, the two-stage hierarchical prior shown in (\ref{Xstage1}) and (\ref{Xstage2}) is a row-sparsity promoting prior which favors most rows of $\bm{X}$ being zeros. 

According to the above hierarchical probabilistic modeling,  we have the joint PDF as
\begin{equation}\label{JointPDF}
	p({\bm{X}},{\bm{Y}},\bm{\alpha},\bm{\beta},\gamma;\mathcal{G}) \!=\! p({\bm{Y}}|{\bm{X}},\bm{\beta}, \gamma;\mathcal{G} )p({\bm{X}}|\bm{\alpha})p(\bm{\alpha} )p(\bm{\beta})p(\gamma).
\end{equation}

\vspace{-0.5cm}
\subsection{Sparse Bayesian Inference}  \label{Section:III:BayesInfer}
Combining the approximation model in subsection (\ref{Section:II:ApproxModel}) and the sparse probabilistic model in  (\ref{Section:III:BayesModel}), we are able to address the  joint optimization subproblem (\ref{iter_subq}) by Bayesian inference. 
In the following, $\!\bm{\Phi}(\mathcal{G},\gamma)$ is abbreviated as $\bm{\Phi}$ for simplicity, and we denote $\bm{A} = \operatorname{diag}(\bm{\alpha})$, $\bm{B} = \operatorname{diag}(\bm{\beta})$.
Bayesian inference starts with the full posterior probability $p({\bm{X}},{\bm{\alpha}},\bm{\beta},\gamma|{\bm{Y}};\mathcal{G})$ which can be decomposed as
\begin{equation}\label{Poster}
	p({\bm{X}},{\bm{\alpha}},\bm{\beta},\gamma|{\bm{Y}};\mathcal{G}) \!=\! p({\bm{X}}|{\bm{Y}},\bm{\alpha} ,\bm{\beta} ,\gamma;\mathcal{G})p(\bm{\alpha},\bm{\beta},\gamma |{\bm{Y}};\mathcal{G}).
\end{equation}
It is shown that the posterior distribution of $\bm{X}$ is Gaussian \cite{BCS_Laplace}
\begin{align}\label{Poster_X}
	p(\bm{X}|{\bm{Y}},\bm{\alpha},\bm{\beta},\gamma;\mathcal{G}) 
	&= \frac{p({\bm{Y}}|{\bm{X}},\bm{\beta},\gamma;\mathcal{G}) p({\bm{X}}|\bm{\alpha})}{p({\bm{Y}}|{\bm{\alpha}},\bm{\beta},\gamma;\mathcal{G})} 
	\notag\\ 
	&=\prod\nolimits_{t=1}^{T}{\cal N}({\bm{x}(t)}|\bm{\mu}(t),\bm{\Sigma}),
\end{align}
with
\vspace{-1em}
\begin{gather}
	\setlength\belowdisplayskip{1pt}
	\bm{\Sigma}= \left(\bm{\Phi}^{{T}}\bm{B} \bm{\Phi} + \bm{A}^{-1} \right)^{-1},\label{Sigma}\\
	\bm{\mu}(t)= \bm{\Sigma} \bm{\Phi} ^{{T}}  {\bm{B }} {\bm{y}}(t) \label{Mu_t}.
\end{gather}

To calculate $\bm{\Sigma}$ and $\bm{\mu}(t)$, we need to estimate the dictionary parameter $\mathcal{G},\gamma$ and  probabilistic model hyperparameters $\bm{\alpha}, \bm{\beta}$. Similar to \cite{RVM2001,ComomSparse,BCS, BCS_Laplace}, type-II maximum likelihood procedure is utilized, thus 
$\bm{\alpha}$, $\beta$, $\mathcal{G}$, and $\gamma$ are approximated by its  maximum a posteriori  probability estimation (MAP). 
\begin{subequations}
	\begin{align}
		( \bm{\alpha}, \bm{\beta}, \mathcal{G},\gamma) &= \mathop {\arg \max }\limits_{\bm{\alpha}, \bm{\beta}, \mathcal{G},\gamma } 
		p(\bm{\alpha},\bm{\beta},\gamma |{\bm{Y}}; \mathcal{G})
		\\
		&= \mathop {\arg \max }\limits_{\bm{\alpha}, \bm{\beta}, \mathcal{G},\gamma} 
		p({\bm{Y}},\bm{\alpha},\bm{\beta},\gamma; \mathcal{G})
		\\
		&= \mathop {\arg \max }\limits_{\bm{\alpha}, \bm{\beta}, \mathcal{G},\gamma} 
		\ln p({\bm{Y}},\bm{\alpha},\bm{\beta},\gamma; \mathcal{G}).\label{paraMAPest}
	\end{align}	
\end{subequations}

In (\ref{paraMAPest}), maximizing the  logarithmic  marginal likelihood $\ln p({\bm{Y}},\bm{\alpha},\bm{\beta},\gamma; \mathcal{G})$ by finding the stationary point is feasible but lacking guaranteed performance since the goal function  $\ln p({\bm{Y}},\bm{\alpha},\bm{\beta},\gamma; \mathcal{G})$ is multimodal and  nonconvex. Instead, we use the  expectation maximization (EM) method to iteratively maximize its evidence lower bound (ELBO)  $ E \left\{
\ln p({\bm{\alpha},\bm{\beta},\bm{Y},\bm{X}},\gamma;\mathcal{G}) \right\}$   by treating $\bm{X}$ as hidden variables,  where $E\left\lbrace \cdot \right\rbrace $ denotes an expectation w.r.t. the posterior of $\bm{X}$ given in (\ref{Poster_X}). As a result,  we have the following update rules.
\vspace{-0.1cm}
\subsubsection{EM Update for Probabilistic Model Parameter $\bm{\beta}$ and $\bm{\alpha}$}
To maximize the ELBO w.r.t $\bm{\beta}$ and $\bm{\alpha}$  is  equivalent to maximize $ E \left\{
\ln p({\bm{X}}|\bm{\alpha})p(\bm{\alpha} )   \right\}$ and  $ E \left\{ p({\bm{Y}}|{\bm{X}},\bm{\beta},\gamma;\mathcal{G}) p(\bm{\beta})   \right\}$ respectively, which leads to the following update rules 
\begin{small}
	%	\vspace{-0.2cm}
	\begin{align}
		\setlength\abovedisplayskip{1pt}
		\setlength\belowdisplayskip{1pt}
		\alpha _i^{new} 	&=\!\frac{\sqrt{ T^2 + 4\lambda \sum\nolimits_{t=1}^{T}{\left( {\Sigma}_{ii} + {\mu}(t)_i^2 \right)}  } -T}{2\lambda}, \text{for } i = 1,\cdots,N, \label{update_alpha}\\
		{\beta_j^{new}} &= \frac{{2a - 2 + T}}{{2b + \sum\nolimits_{t=1}^{T}\left( \text{Res}(t)_j^2 + \Delta_{jj} \right)   }},\text{for } j = 1,\cdots,M,  \label{update_beta}
	\end{align}
\end{small}

\noindent with $\textbf{Res}(t) = \bm{y}(t) - \bm{\Phi}\bm{\mu}(t)$, $\bm{\Delta} = \bm{\Phi} \bm{\Sigma} \bm{\Phi}^T$. For simplicity expression, the derivations of (\ref{update_alpha}) and  (\ref{update_beta})  are presented in Appendix \ref{AppendixC} and  Appendix \ref{AppendixD}, respectively.

\vspace{0.1cm}	
\subsubsection{EM Update for Dictionary Model Parameter  $\mathcal{G}$ and $\gamma$}
The localization dictionary is parametrized by $\mathcal{G}$ and $\gamma$, thus to learn the sparsifying localization dictionary  is equal to learn the corresponding dictionary parameters. According to (\ref{JointPDF}), the maximization of the ELBO w.r.t. to  $\mathcal{G}$ and $\gamma$ is equivalent to maximize $E\left\{ \ln  p({\bm{Y}}|{\bm{X}},\bm{\beta},\gamma;\mathcal{G} ) p(\gamma) \right\}$  which is tantamount to minimize 
{\small
	\begin{align}\label{E_step_theta}
		\setlength\abovedisplayskip{1pt}
		\setlength\belowdisplayskip{1pt}
		&E\left\{ \sum_{t=1}^{T} \left( \bm{y}(t) - \bm{\Phi} \bm{x}(t)\right)^{T} \bm{B} \left( \bm{y}(t) - \bm{\Phi} \bm{x}(t)\right) \right\} 
		\notag\\
		&=\!\! \sum\limits_{t = 1}^T \!\!\Bigg\{\left( \bm{y}(t) - \bm{\Phi} \bm{\mu}(t)\right)^{T} \bm{B} \left( \bm{y}(t) - \bm{\Phi} \bm{\mu}(t)\right) + \operatorname{tr}\left(\bm{\Phi} \bm{\Sigma}\bm{\Phi} ^{T} \bm{B}\right) \!\Bigg\}.
\end{align}}

%\vspace{0.1cm}
By incorporating the dictionary approximation model  (\ref{iter_subq:c1}),( \ref{iter_subq:c2}) into above goal function, minimizing (\ref{E_step_theta}) boils down to solving the following linear least square (LLSQ) problem with boundary constraints as 
\begin{subequations} \label{delta_LLSQ}
	\begin{gather}
		\mathop {\arg \min }\limits_{ \bm{\delta_{u}}, \bm{\delta_{v}}, \delta_{\gamma} }  
		\left\{ \begin{array}{l}
			\bm{\delta}_u^T  {\bm{M}_{uu}} 	\bm{\delta}_u + \bm{\delta}_v^T  {\bm{M}_{vv}} 	\bm{\delta}_v  + p\delta _\gamma ^2 + 2 \bm{\delta}_u^T  {\bm{M}_{uv}}  {\bm{\delta} _v}\\
			+ 2{\delta _\gamma } \bm{v}_{u\gamma }^T  {\bm{\delta} _u} + 2{\delta _\gamma }\bm{v}_{v\gamma }^T  {\bm{\delta}_v} + 2\bm{v}_u^T{\bm{\delta}_u} + 2\bm{v}_v^T{\bm{\delta}_v} + 2q{\delta _\gamma }
		\end{array} \right\}\label{delta_LLSQ:goal}\\
		\text{s.t.}\;  \bm{\delta}_{u} \in [\bm{LB}_u, \bm{UB}_u], \bm{\delta}_{v} \in [\bm{LB}_v,\bm{UB}_v], \delta_{\gamma} \in [LB_{\gamma}, UB_{\gamma}]. \label{boundary}
	\end{gather}
\end{subequations} 
with
\vspace{-1em}
\begin{subequations} 
	\begin{gather}
		\bm{M}_{uu} = {\bm{\Phi}'_{u}}^T \bm{B} {\bm{\Phi}'_{u}}  \circ \left( T \cdot \bm{\Sigma}  + \bm{U} \bm{U}^{T }  \right),\\
		\bm{M}_{vv} = {\bm{\Phi}'_{v}}^T \bm{B} {\bm{\Phi}'_{v}}  \circ \left( T \cdot \bm{\Sigma}  + \bm{U} \bm{U}^{T }  \right),\\
		\bm{M}_{uv} = {\bm{\Phi}'_{u}}^T \bm{B} {\bm{\Phi}'_{v}}  \circ \left( T \cdot \bm{\Sigma}  + \bm{U} \bm{U}^{T }  \right),\\
		\bm{v}_{u\gamma } = \left[ {\bm{\Phi}'_{u}}^T \bm{B} {\bm{\Phi}'_{\gamma}}  \circ \left( T \cdot \bm{\Sigma}  + \bm{U} \bm{U}^{T }  \right) \right] \cdot \bm{1}_N,\\
		\bm{v}_{v\gamma } = \left[ {\bm{\Phi}'_{v}}^T \bm{B} {\bm{\Phi}'_{\gamma}}  \circ \left( T \cdot \bm{\Sigma}  + \bm{U} \bm{U}^{T }  \right) \right] \cdot \bm{1}_N,\\
		{\small \bm{v}_{u} \!=\!  T  \cdot \operatorname{diag}{\left( {\bm{\Phi}'_{u}}^T \bm{B} \bm{\Phi}_0 \bm{\Sigma} \right)}
			\!-\! \sum\limits_{t = 1}^T \operatorname{diag}\Big( {\bm{\mu}(t)}\Big) {\bm{\Phi}'_{u}}^T \bm{B} \Big( {\bm{y}(t) \!-\!  \bm{\Phi}_0 \bm{\mu}(t) } \Big)},\\
		{\small \bm{v}_{v} \!=\!  T  \cdot \operatorname{diag}{\left( {\bm{\Phi}'_{v}}^T \bm{B} \bm{\Phi}_0 \bm{\Sigma} \right)}
			\!-\! \sum\limits_{t = 1}^T \operatorname{diag}\Big( {\bm{\mu}(t)}\Big) {\bm{\Phi}'_{v}}^T \bm{B} \Big( {\bm{y}(t) \!-\!  \bm{\Phi}_0 \bm{\mu}(t) } \Big)},\\
		p = T \cdot \operatorname{tr}\left\{   {\bm{\Phi}'_{\gamma}} \Sigma  {\bm{\Phi}'_{\gamma}}^T  \bm{B} \right\} +  \operatorname{tr}\left\{ \bm{U}^{T}  {\bm{\Phi}'_{\gamma}}^T \bm{B} {\bm{\Phi}'_{\gamma}} \bm{U} \right\} \\
		%\sum_{t=1}^{T}\bm{\mu}(t)^T {\bm{\Phi}'_{\gamma}}^T \bm{B} {\bm{\Phi}'_{\gamma}}\bm{\mu}(t),
		q = T \cdot \operatorname{tr}\left\{   \bm{\Phi}_0 \Sigma   {\bm{\Phi}'_{\gamma}}^T \bm{B} \right\} - \operatorname{tr}\left\{ \left( \bm{Y} -  \bm{\Phi}_0 \bm{U}  \right)^T  \bm{B} {\bm{\Phi}'_{\gamma}} \bm{U} \right\},
	\end{gather}
\end{subequations} 

\noindent and $\bm{U}= \left[\bm{\mu}(t),\cdots, \bm{\mu}(t) \right]$. In (\ref{boundary}), $\bm{\delta}_{u}$, $\bm{\delta}_{v}$ and ${\delta}_{u}$ are bounded for the fact that the first-order approximation is only valid in the vicinity of the expansion point. 	For simplicity expression, the detailed derivations of (\ref{E_step_theta}) and  (\ref{delta_LLSQ})  are presented in Appendix \ref{AppendixA} and  Appendix \ref{AppendixB}, respectively.

Denote by $f(\bm{\delta})$ the goal function in (\ref{delta_LLSQ:goal}), which is convex  and can be globally minimized using a variety of standard optimization packages. we provide here an analytical solution for
problem (\ref{delta_LLSQ}) as follows. First, we have the partial derivative with respect to $\bm{\delta}_u$ as
\begin{equation}
	\frac{\partial f}{\partial \bm{\delta} _u} = 2\left( {\bm{M}_{uu}}    {\bm{\delta} _u} + {\bm{M}_{uv}}  {\bm{\delta}_v} + {\delta _\gamma }{\bm{v}_{u\gamma }} + {\bm{v}_u} \right).
\end{equation}
Thus, the minimum is achieved at {\small $\bm{\delta}_u^{\ast}  =  -\bm{M}_{uu}^{-1}({\bm{M}_{uv}}  {\bm{\delta}_v} + {\delta _\gamma }{\bm{v}_{u\gamma }} + {\bm{v}_u} )$} if ${ \bm{M}_{uu} }$ is invertible and {\small ${ \bm{\delta}^{\ast}_{u} \!\in\! \left[ \bm{LB}_{u}, \bm{UB}_{u} \right]} $}. Then, we have $\bm{\delta}^{l+1}_u =\bm{\delta}^{\ast}_{u} $. Otherwise, we update $\bm{\delta}_{u}$ element by element. 
Fix other elements but  $({\delta}_{u})_i$, denote by  $(\bm{\delta}_{u})_{-i}$ the vector $\bm{\delta}_{u}$ without the $i$-th entry $({\delta}_{u})_i$,  and then the solution to the $i$-th stationary point equation of {\small $ (\bm{M}_{uu}^i )_{-i}  (\bm{\delta}_u )_{-i}  +  ( M_{uu} )_{ii}  (\delta_u )_{i} + ( {\bm{M}_{uv}}  {\bm{\delta}_v} + {\delta _\gamma }{\bm{v}_{u\gamma }} + {\bm{v}_u} )_i  =0 $} is
\begin{equation}
	\setlength\abovedisplayskip{2pt}
	\setlength\belowdisplayskip{2pt}
	({\tilde\delta}_{u})_{i}  = -\frac{ (\bm{M}_{uu}^i )_{-i}  (\bm{\delta}_u )_{-i}   + ( {\bm{M}_{uv}}  {\bm{\delta}_v} + {\delta _\gamma }{\bm{v}_{u\gamma }} + {\bm{v}_u} )_i }{  ( M_{uu} )_{ii} }.
\end{equation}

Therefore, the elementwise update of ${\bm \delta}^{\l+1}_{u}$ is 
\begin{equation}
	\setlength\abovedisplayskip{1pt}
	\setlength\belowdisplayskip{1pt}
	\left(  {\delta}^{\l+1}_{u} \right)_i =  
	\begin{cases}
		(LB_{u} )_i, & \text{if~} ({\tilde\delta}_{u})_{i} < (LB_{u} )_i;\\
		({\tilde\delta}_{u})_{i}, &\text{if~} ({\tilde\delta}_{u})_{i} \in { \left[ (LB_{u} )_i, ( UB_{u} )_i  \right] };  \\
		(UB_{u} )_i, &{ \text{otherwise.}}
	\end{cases}	
\end{equation}

Following  the same steps, we have the update rule for $\bm{\delta}_v$ as
$ \bm{\delta}_v^{l+1} = \bm{\delta}_v^{\ast}  =  -\bm{M}_{vv}^{-1}( {\bm{M}_{uv}^T}  {\bm{\delta}_u} + {\delta _\gamma }{\bm{v}_{v\gamma }} + {\bm{v}_v} )$ if ${ \bm{M}_{vv}}$ is invertible and {\small ${ \bm{\delta}^{\ast}_{v} \!\in\! \left[ \bm{LB}_{v}, \bm{UB}_{v} \right]} $}. Otherwise, we have elementwise update as
\begin{equation}
	\setlength\abovedisplayskip{1pt}
	\setlength\belowdisplayskip{1pt}
	\left(  {\delta}^{\l+1}_{v} \right)_i =  
	\begin{cases}
		(LB_{v} )_i, & \text{if~} ({\tilde\delta}_{v})_{i} < (LB_{v} )_i;\\
		({\tilde\delta}_{v})_{i}, &\text{if~} ({\tilde\delta}_{v})_{i} \in { \left[ (LB_{v} )_i, ( UB_{v} )_i  \right] };  \\
		(UB_{v} )_i, &{ \text{otherwise.}}
	\end{cases}	
\end{equation}
with 
\begin{equation}
	\setlength\abovedisplayskip{2pt}
	\setlength\belowdisplayskip{2pt}
	({\tilde\delta}_{v})_{i}  = -\frac{ (\bm{M}_{vv}^i )_{-i}  (\bm{\delta}_v )_{-i}   + ( {\bm{M}_{uv}^T}  {\bm{\delta}_u} + {\delta _\gamma }{\bm{v}_{v\gamma }} + {\bm{v}_v} )_i }  {  ( M_{vv} )_{ii} }.
\end{equation}

In particular, minimizing problem (\ref{delta_LLSQ}) with respect to $\delta_{\gamma}$ degenerates to a scalar quadratic function  optimization problem 
\begin{equation}\label{delta_gamma_qf}
	\setlength\abovedisplayskip{1pt}
	\setlength\belowdisplayskip{3pt}
	\mathop {\arg \min }\limits_{  {{\delta}_{\gamma}} \in \left[   LB_\gamma, UB_\gamma \right] } 
	\Big\{
	p{\delta}_{\gamma}^{2} +  2\left(  \bm{v}_{u\gamma }^T  {\bm{\delta} _u} +\bm{v}_{v\gamma }^T  {\bm{\delta}_v} + q \right){\delta}_{\gamma} 
	\Big\}.
\end{equation}

Note that $p>0$, hence its minimum can be achieved either at the boundary ($LB_\gamma$ or $UB_\gamma$) or at the axis of symmetry 
\begin{equation}
	\setlength\abovedisplayskip{2pt}
	\setlength\belowdisplayskip{2pt}
	\delta_{\gamma}^{\ast} = - \frac{\bm{v}_{u\gamma }^T  {\bm{\delta} _u} +\bm{v}_{v\gamma }^T  {\bm{\delta}_v} + q}{p}.
\end{equation}

\vspace{-1em}
\subsection{The Proposed PSBDL Algorithm}  \label{Subsection:DEMSL}
Based on the above analysis, the proposed parametric sparse Bayesian dictionary learning (PSBDL) algorithm is summarized in Algorithm \ref{Alg:PSBDL}. According to the outputs of Algorithm \ref{Alg:PSBDL}, we retrieve the source locations, the transmitted powers as follows.

\subsubsection{Source locations  and transmitted powers estimation}
As in traditional CS-based approaches, we can estimate the spatial power spectrum of the sources with $\hat{\bm{X}}$ and $\hat{\mathcal{G}}$. 	
Recall the probabilistic modeling in (\ref{HPM}), for each row $\bm{ X}^i$ in the posterior estimation of $\bm{X}$, we have  $\bm{ X}^i \sim  \mathcal{N}(\bm{U}^i, {\Sigma}_{ii}\bm{I})$. Thus, the expectation of the spatial power  spectrum  strength at GP $\hat{\bm{g}}_i$ is
{\small
	\begin{equation}\label{exp_power}
		\setlength\abovedisplayskip{2pt}
		\setlength\belowdisplayskip{2pt}
		{\hat{P}_i} = E\left\{\frac{\bm{ X}^i \bm{1}_T}{T}\right\}  = \frac{\bm{ U}^i \bm{1}_T}{T}= \frac{\hat{\bm{ X}}^i \bm{1}_T}{T}, \text{~for~} i= 1,\cdots,N. 
	\end{equation}
}
\noindent Then, the source locations are estimated by the GPs with the highest $K$ peaks of the spatial power spectrum, the transmitted powers are estimated with the expected spatial power  spectrum  strength of the corresponding GPs.

\subsubsection{Implementation details and computational complexity}
First, $N$ is artificially decided candidate grid point number, which is prone to be a large number greater than $M$. Thus, the matrix inversion operation for calculating $\bm{\Sigma}\!\in\! \mathbb{R}^{N \times N}$ in Step (7) is complex and time-consuming. To this end,  Woodbury matrix identity is used, when $N>M$, to reduce  the dimension of the matrix inversion from $N$ to $M$  
\begin{equation}
	\setlength\abovedisplayskip{2pt}
	\setlength\belowdisplayskip{2pt}
	\bm{\Sigma} \!=\!  \bm{A} \!-\!  \bm{A} {\bm{\Phi}^T} \bm{\Xi}^{-1} {\bm{\Phi}}  \bm{A}
\end{equation}
where $\bm{\Xi}= \bm{B}^{-1} + \bm{\Phi} \bm{A} \bm{\Phi}^{T}$. As such,  the matrix inversion computation complexity is reduced from ${\mathcal O}(N^3)$ to ${\mathcal O}(M^3)$  since $\bm{\Xi}\!\in\! \mathbb{R}^{M \times M}$.

Furthermore, the dictionary approximation in Step (6) and reconstruction in Step (4) can be constrained to these grid points where the rows of $\bm{X}$ are non-zero. In the proposed probabilistic model, these non-zero rows are characterized by greater variance $\alpha_i$. As a result, we infer ${\bm \delta}_{u}$ and ${\bm \delta}_{v}$ of the GPs  with the highest $K$ variances. In this way, the coefficient matrix and coefficient vector in the goal function $f(\bm{\delta})$ can be truncated into dimension $K \times K$ or  $K \times 1$, which is crucial to reduce the LLSQ problem dimension from $N+1$ to $K+1$ and thus to speed up the algorithm.

Based on the above implementation details, the computational complexities per iteration for the main steps are: $\mathcal{O}(KM)$ for Step (4);  $\mathcal{O}(MK^2)$ for Step (6); $\mathcal{O}(MN^2 + M^{2}N + M^{3})$ for computing $\bm{\Sigma}$ and $\mathcal{O}(TMN^2)$ for computing $U$ in Step (7); $\mathcal{O}(N)$ for Step (9). Generally we have $ K \!<\! M \!<\! N$, thus the asymptotic complexity per  iteration for the proposed PSBDL algorithm is $\mathcal{O}(TMN^2)$.

\setlength{\textfloatsep}{1pt}% Remove \textfloatsep
\begin{algorithm}[t]\label{Alg:PSBDL}
	%	footnotesize
	%	\SetAlgoNoLine
	%\emph{// Grid evolution}\\
	\begin{small}
		\caption{\small Parametric Sparse Bayesian Dictionary Learning }
		\KwIn{$\bm{Y}$, $K$, $N$, $\mathcal{S}$, propagation model $f(\bm{s}_i,\bm{t}_k,\gamma )$.}
		\KwOut{$\bm{\hat X}$,  $\hat{\mathcal{G}}$ and $\hat{\gamma}$;}	
		Initialize $\mathcal{G}=\mathcal{G}^{(0)}$, $\gamma = \gamma^{(0)}$, $\bm{\alpha}$, $\bm{\beta}$, $\lambda$, $a$, $b$, $k=0$\;
		\While{external loop stopping condition not hold }{	
			$\bm{\delta}^0_u=\bm{0}$, $\bm{\delta}^0_v=\bm{0}$, $\delta^0_{\gamma} = 0$,  $l=0$\;
			\tcp{\footnotesize Dictionary update}
			Calculate $\bm{\Phi_0}$, $\bm{\Phi}^{\prime}_{u}$, $\bm{\Phi}^{\prime}_{v}$, $\bm{\Phi}^{\prime}_{\gamma}$ using $\mathcal{G}^{(k)}$ and $\gamma^{(k)}$\;
			\While{internal loop stopping condition not hold }{
				\tcp{ \footnotesize Sparse recovery and dictionary approximation} 
				Update $\bm{\Phi}$ by  (\ref{iter_subq:c2}) using $\bm{\delta}_u^{(l)}$, $\bm{\delta}_v^{(l)}$, $\delta_{\gamma}^{(l)}$\; 
				Compute $\bm{\Sigma}$ and $\bm{U}$ using $\bm{\alpha}^l$, $\bm{\beta}^l$ and $\bm{\Phi}$\;
				Update $\bm{\alpha}^{l+1}$, $\bm{\beta}^{l+1}$ according to (\ref{update_beta}), (\ref{update_alpha})\; 
				Calculate $\bm{\delta}_u^{(l+1)}$, $\bm{\delta}_v^{(l+1)}$,${\delta}_{\gamma}^{(l+1)}$ by solving
				({\ref{delta_LLSQ}})  \;   
				$l = l+1$ ; //{\sl Update internal loop iteration counter}   
			}
			$\mathcal{G}^{(k+1)} =  \mathcal{G}^{(k)} + {\bm{\delta}}_g^{(l)}$,  $\gamma^{(k+1)} = \gamma^{(k)} + {\delta}_{\gamma}^{(l)}$\;
			$k = k+1$; //{\sl Update external loop iteration counter} 
		}
		\KwRet{ $\bm{\hat X} =  \bm{ U }$, $\hat{\mathcal{G}}= \mathcal{G}$, $\hat{\gamma}= \gamma$; }
	\end{small}
\end{algorithm}

\vspace{-0.2cm}
\section{ Cram{\'e}r-Rao Bound Analysis}
\vspace{-0.1cm}

In this section, we 
%	assume the transmitted powers are constant during the MMV measurement period, and
derive the Cram{\'e}r-Rao lower bound (CRLB) as an estimation benchmark for the unknown parameter vector {\small$\bm{\vartheta} \!=\! \left[ u^t_1, v^t_1, P_1, \cdots, u^t_K, v^t_K, P_K, \gamma ,\beta_1,\cdots,\beta_M\right]^T $}. In estimation theory, the CRLB provides a theoretical performance  limit for any unbiased estimator of the source locations $[u^t_k, v^t_k]$, the transmitted powers $P_k$ for $k \!=\! 1,\!\cdots \!, K$, as well as the PLE $\gamma$ in the presence of unknown nonuniform Gaussian noise variance $\beta_1,\cdots,\beta_M$, given the observation $Y$.  

Indeed, the CRLB gives a lower bound for the error covariance matrix
\vspace{-0.2cm}
\begin{equation}
	\setlength\abovedisplayskip{4pt}
	\setlength\belowdisplayskip{2pt}
	E\left\{ (\bm{\hat \vartheta} - \bm{ \vartheta})(\bm{\hat \vartheta} - \bm{ \vartheta})^T\right\} \ge \bm{J}^{-1},
\end{equation}
where the inequality sign is defined in the positive-semidefinite (PSD) sense. $\bm{J} \!\in\! \mathbb{R}^{(3K+M+1) \times (3K+M+1)}$ is the Fisher information matrix (FIM) defined as 
\begin{equation}
	\bm{J}   =  E \left\{ -\Delta_{\bm{\vartheta}}^{\bm{\vartheta}} \ln p\left(\bm{Y};\bm{\vartheta}  \right)  \right\}
\end{equation}
with $\Delta_{\bm{\vartheta}}^{\bm{\vartheta}} \!\!=\!\! \nabla_{\bm{\vartheta}}  \nabla_{\bm{\vartheta}} ^T $ being the second derivative (Hessian) operator, and $\nabla_{\bm{\vartheta}} $ being the gradient operator with respect to $\bm{\vartheta}$.

Using the Gaussian observation model in and considering the MMV case,  we have 
\begin{equation}
	\bm{J}  = \sum\limits_{t=1}^T  \underbrace {E \left\{ -\Delta_{\bm{\vartheta}}^{\bm{\vartheta}} \ln p\left(  \bm{y}(t);\bm{\vartheta}  \right)  \right\} }_{\bm{J}_t}.
\end{equation}
with $\bm{J}_t  \!\in\! \mathbb{R}^{ (3K+M+1) \times (3K+M+1)} $ being the FIM of the  snapshot measurement $\bm{y}(t)$. The PDF of  each $\bm{y}(t)$ is   
\begin{equation}
	p\left(\bm{y}(t);\bm{\vartheta}  \right) = {\mathcal{N}}({\bm{y}(t)}|\bm{\tilde{\mu}}, \bm{B}^{-1})
\end{equation}
with  $\tilde{\mu}_i \!=\! \sum \nolimits_{ k = 1}^K {P_k} {f(\bm{s}_i,\bm{t}_k,\gamma)}$, for $i \!=\! 1,\cdots, M$ and $\bm{B} = \operatorname{diag}(\beta_1,\cdots,\beta_M)$. For the Gaussian observation vector $\bm{y}(t)$, the $(i,j)$-th element of the  FIM $\bm{J}_t$ can be computed as \cite[Ch. 3]{kay1993fundamentals}
\begin{equation} \label{FIM_yt}
	[\bm{J}_t]_{i, j}=\frac{\partial \tilde{\bm\mu}^{T}}{\partial \vartheta_{i}} \bm{B}\frac{\partial \tilde{\bm\mu}}{\partial \vartheta_{j}} +\frac{1}{2} \operatorname{tr}\left(\bm{B} \frac{\partial \bm{B}^{-1}}{\partial \vartheta_{i}} \bm{B} \frac{\partial \bm{B}^{-1}}{\partial \vartheta_{j}}\right) 
\end{equation}	
with 
\begin{subequations} \label{partials}
	\begin{gather}
		\frac{\partial \tilde{\bm\mu}}{\partial \vartheta_{i}} =\left[ \frac{\partial {\tilde\mu_1}}{\partial \vartheta_{i}},\cdots, \frac{\partial {\tilde\mu_M}}{\partial \vartheta_{i}} \right]^{T}, \\
		\frac{\partial \bm{B}^{-1}}{\partial \vartheta_{i}} = \operatorname{diag}\left( \frac{\partial {\beta_1^{-1}}}{\partial \vartheta_{i}},\cdots, \frac{\partial {\beta_M^{-1}}}{\partial \vartheta_{i}} \right).
	\end{gather}
\end{subequations}

For example, if the path loss model is ${f(\bm{s}_i,\bm{t}_k,\gamma )}= \left( d_{ik}\right)^{-\gamma}$ with $d_{ik} = \lVert \bm{s}_i -\bm{t}_k \rVert_2$, then for $k=1,\cdots, K$ and  $i,j = 1,\cdots,M$, the non-zero partial derivative terms in (\ref{partials}) are
\begin{subequations}
	\begin{gather}
		\frac{\partial {\tilde \mu}_i}{\partial u^t_k} =P_k \frac{\partial {f(\bm{s}_i,\bm{t}_k,\gamma)} } {\partial u^t_k} = -\gamma P_k \frac{ u^t_k -  u^s_i}{d_{ik}^{\gamma +2}}, \label{CRB1}\\
		\frac{\partial {\tilde \mu}_i}{\partial v^t_k} =P_k \frac{\partial {f(\bm{s}_i,\bm{t}_k,\gamma)} } {\partial v^t_k} = -\gamma P_k \frac{ v^t_k -  v^s_i}{d_{ik}^{\gamma +2}}, \label{CRB2}\\
		\frac{\partial {\tilde \mu}_i}{\partial P_k}   =  f(\bm{s}_i,\bm{t}_k,\gamma) =  \frac{1}{ d_{ik}^{\gamma} }.  \label{CRB3}\\
		\frac{\partial {\tilde \mu}_i}{\partial \gamma} \!=\! \frac{\partial   \sum \nolimits_{ k = 1}^K {P_k} {f(\bm{s}_i,\bm{t}_k,\gamma)} } {\partial v^t_k} \!=\! -\! \! \sum \limits_{k = 1}^K  { \! P_k \frac{\ln d_{ik}}{d_{ik}^{\gamma}}     }, \label{CRB4}\\
		\frac{\partial \beta_j^{-1}}{\partial \beta_j} = - \frac{1}{\beta_j^2}.
	\end{gather}         
\end{subequations}

Generally, we tend to express the powers in  decibels ${P_k}_{dB} = 10 \lg P_k$, thus the  partial derivative with respect to powers in decibels is 
\begin{equation}
	\frac{  \partial {\tilde \mu}_i  }{  \partial {P_k}_{dB} } = \frac{ \partial {\tilde \mu}_i }{ \partial P_k }  \frac{ \partial P_k }{\partial {P_k}_{dB}}= \frac{P_k \ln 10}{10 d_{ik}^{\gamma} }.
\end{equation}

%	Finally, based on the above analysis,  we complete this section with the following theorem.
%	\begin{theorem}
%		For the multiple sources localization problem with unknown PLE, the FIM $\bm{J}$, the inverse of the CRLB matrix, is
%		\begin{equation}
%			\bm{J} = T \cdot    \left[ \frac{\partial \bm{\tilde{\mu}} }{ \partial \bm{\vartheta}^T } \right]^T \bm{B} \left[ \frac{\partial \bm{\tilde{\mu}} }{ \partial \bm{\vartheta}^T } \right]
%		\end{equation}
%		where the entries of the  Jacobian matrix  ${\partial \bm{\tilde{\mu}} }/{ \partial \bm{\vartheta}^T }$ is given in (\ref{JacobMtx}).
%	\end{theorem}

\section{Numerical Simulations}
In this section, we evaluate the localization performance of the proposed method by numerical simulations. The simulation setup refers to \cite{Gholami2} where a square localization area of 20 m by 20 m is considered,  and the path loss model is set to  ${f(\bm{s}_i,\bm{t}_k,\gamma )}[dB] \!=\! -10{\gamma}\lg\lVert \bm{s}_i -\bm{t}_k \rVert_2$ when $\lVert \bm{s}_i -\bm{t}_k \rVert_2 > 1 \text{m}$, and ${f(\bm{s}_i,\bm{t}_k,\gamma )}[dB] \!=\! 0$ otherwise. Like  \cite{Gholami2}, the path-loss exponent $\gamma$ is randomly drawn from [2,6], and the transmitted powers are randomly drawn from [-10 dBm, 0 dBm] in each trail.  Different from \cite{Gholami2} where only one source is considered and the sensors' locations are fixed in each trail,  we deploy three sources at  [5, 9], [11,17] and [15, 5], all in meters and randomly deploy the sensors  inside the area in each trail to avoid sticking to any specific sensor network topology. 	
The simulation results  are averaged over $N_{s} \!=\!500$ randomized trials  carried out in Matlab R2016a on a PC with Windows 10 OS and an Intel i7-6700 CPU.

In the simulation,  the nonuniform noise is modeled as $\varepsilon_{i}(t)\sim \mathcal{N}\left(0, \sigma^{2}_i\right)$ with $  \sigma_1 \neq \sigma_2 \neq,  \cdots,\neq \sigma_M $, for which we define the signal-to-noise ratio (SNR) of the $i$-th sensor as $10\lg\left( \lVert \left(\bm{\Phi}(\mathcal{T}, \gamma) \bm{W}\right) ^{i}\rVert_2^2/\left(T\sigma_i^2\right)\right)$.
%	 Note that in each randomized trail, we assume the transmitted powers are constant across all the snapshots. 
The evaluation metric is the  root-mean-square error (RMSE) defined as 
\begin{equation}
	\text{RMSE} =  \sqrt{\frac{1}{K}\sum_{k=1}^{K}\lVert \bm{\theta}_k - \bm{\hat{\theta}}_k\rVert^2_2} 
\end{equation} 
with $\bm{\hat{\theta}}_k$ being the $k$-th estimated parameter of truth $\bm{{\theta}}_k$, where $\bm{\theta}_k$ denotes $\bm{t}_k$ for source locations estimation, $P_k$ for source powers estimation, and $\gamma$ for path-loss exponent estimation with $K=1$, respectively.

%    Moreover, assume that the maximum localization error at the $n$-th randomized trail, i.e.,
%    \begin{equation}
%      \delta^{n}_{\text{max}} := \mathop{\max}\limits_{k\in \left\{ 1,\cdots,K \right\} } {\lVert \bm{t}^n_k - \bm{\hat{t}}^n_k\rVert_2}, 
%    \end{equation}
%    is a draw of the random variable $\Delta$. We further define the localization error function $P_d$ as
%    \begin{equation}
%    	P_{d} := \Pr\left( \Delta > d\right) = 1- F_{\Delta}(d),
%    \end{equation}
%   which represents the probability that at least one source is localized with an error of more than $d$ meters. Here, $ F_{\Delta}(d)$ is the empirical cumulative distribution function (cdf) of the error $\Delta$. 
%    

To investigate the effectiveness of the proposed approach, we compare with the traditional sparsity-based MSL algorithm M-SBL\cite{MSBL},  the state-of-the-art off-grid MSL algorithm GEMTL\cite{GEMTL}, as well as the theoretical limits CRLB derived in Section IV. Both M-SBL and GEMTL assume the measurement noise to be uniform, and specifically, M-SBL uses the initialized parameters to form a fixed localization dictionary, while GEMTL partly infers the localization dictionary by modeling the off-grid offsets.   Note that the original  GEMTL algorithm is designed for SMV case, hence in the simulation, we extend it to the MMV case and term it as M-GEMTL. 
The path-loss exponent  $\gamma$ is initialized as 2 for all algorithms, and the set of candidate GPs is initialized with a uniform grid points. We examine the performance from different aspects shown as follows.

\subsection{Impacts of Different Grid Granularity}

First of all, since the number of GPs is an artificially decided parameter which may affect the estimation performance,  in this simulation, we set SNR = 25dB, $M$ = 60, and $T$ = 5 to study the impacts of  the grid granularity defined as $\sqrt{N}$ with $N$ being the number of candidate GPs. It is worth noting that the CRLB is constant for all granularity since it is irrelative to the grid discretization.

\begin{table}[t]
	\newcommand{\tabincell}[2]{\begin{tabular}{@{}#1@{}}#2\end{tabular}}
	\centering
	%\setlength{\abovecaptionskip}{0pt}%
	%\setlength{\belowcaptionskip}{10pt}%
	%	\caption{\label{tab:11}The performance of four graph learning methods from signals generated from a random geometric graph.}
	\caption{RMSE of PLE estimate versus the grid granularity }
	\vspace{-0.2cm}
	%\Xhline{1.2pt}
	%\setlength{\abovecaptionskip}{0pt}
	%\setlength{\belowcaptionskip}{10pt}
	\begin{tabular}{ c c c c c  c }
		%\Xhline{1.2pt}
		\toprule
		%\hline
		%\specialrule{0em}{2pt}{2pt}
		%\tabincell{c}
		\textbf{Grid granularity}  &  \tabincell{c}{6} & \tabincell{c}{8 } & \tabincell{c}{10 } &\tabincell{c}{12}  & \tabincell{c}{14} \\
		\midrule
		\textbf{CRLB}    &0.0046    &0.0045    &0.0045    &0.0045    &0.0046    \\
		%40 &0.7154   &0.4678   &0.8280   &2.0313  & 9.2731    \\
		\textbf{PSBDL}   &0.1211    &0.0814    &0.0662    &0.0661    &0.0661   \\
		\bottomrule	
	\end{tabular}
	\label{tab:PLE_granularity}
\end{table}

\begin{figure}[t]
	\setlength{\abovecaptionskip}{-0.05cm}
	\begin{minipage}[b]{0.49\linewidth}
		\centerline{\includegraphics[height = 7.5cm]{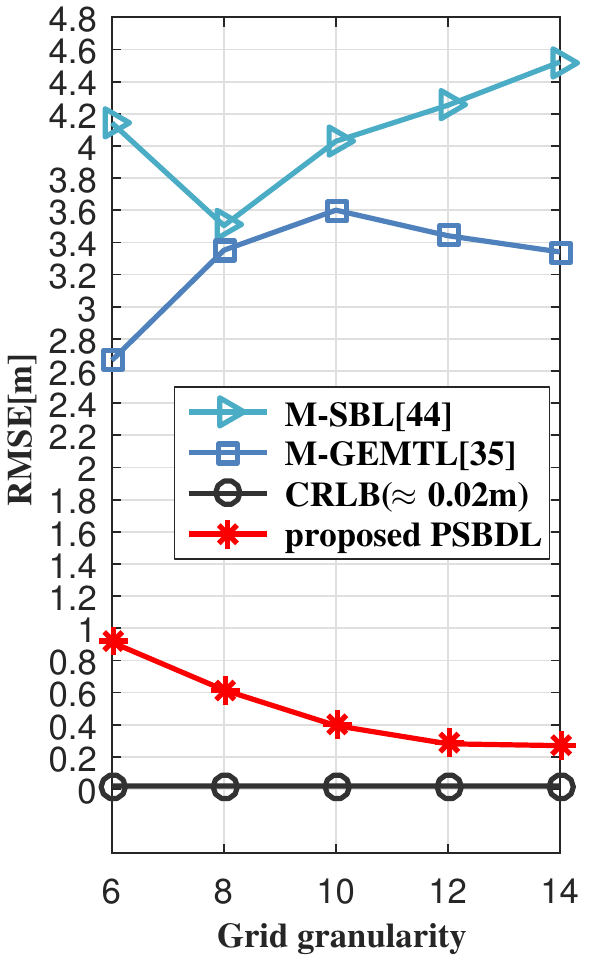}}
		\vspace{-0.1cm}
		\centerline{(a)}
	\end{minipage}
	\begin{minipage}[b]{0.49\linewidth}
	   \centerline{\includegraphics[height = 7.5cm]{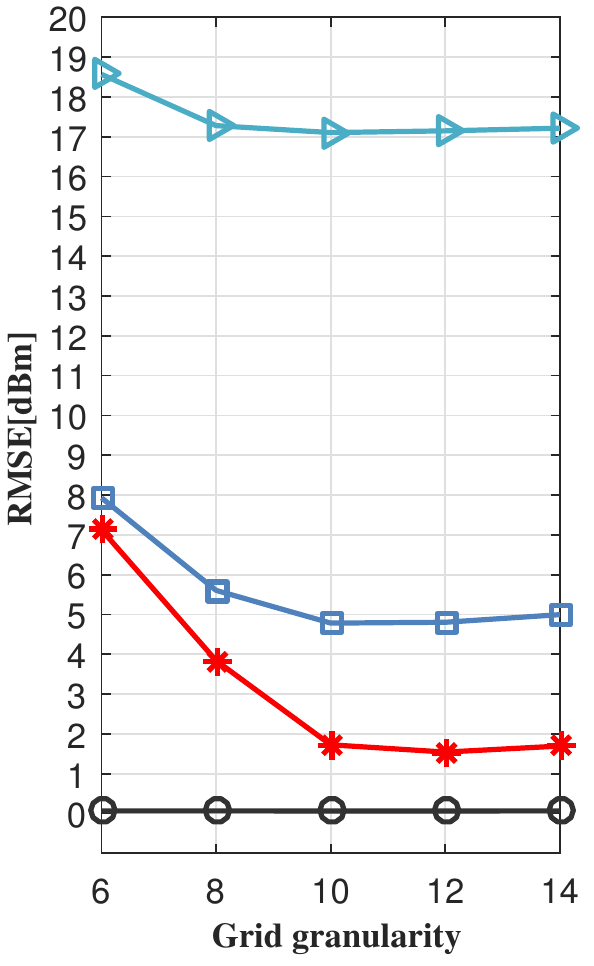}}
		\vspace{-0.1cm}
		\centerline{(b)}
		%\medskip
	\end{minipage}
	\caption{The RMSE of different approaches versus the grid granularity for (a) the locations estimate and (b) the transmitted powers estimate.  }
	\label{fig:grid_granularity}
	%	\vspace{-0.4cm}
\end{figure}

Tab. \ref{tab:PLE_granularity} presents the RMSE of the PLE estimate for the  proposed method when the grid granularity changed from 6 to 14. 
It is observed that as the grid granularity increases, the RMSE of the proposed PSBDL algorithm reduces from 0.1211 to 0.0061 and gradually approaches the CRLB. Specifically, when the grid granularity is less than 10, as the grid granularity increases the RMSE decreases, which is because the finer grid granularity, the higher possibility to capture the off-grid sources, the higher possibility to alleviate the dictionary mismatch, and the higher accuracy of the PLE estimates. When the grid granularity is greater than 10, the RMSE of PLE almost remains constant, which is because the granularity of 10 is enough for the proposed method to capture the off-grid sources, thus the grid granularity is no longer a major influencing factor. 

Fig. \ref{fig:grid_granularity} illustrates the RMSE  of the  locations and transmitted powers estimates for different approaches when grid granularity changes.  As expected, benefiting from the inference of the propagation parameters and the proper candidate GPs, the proposed method exhibits the lowest estimation errors, and its RMSE  decreases as the grid granularity increases and is very close to the CRLB. Similar to Tab. \ref{tab:PLE_granularity}, when the grid granularity greater than 12, the RMSE of location estimation and transmitted powers tend to be converged.
In contrast, for M-SBL and M-GEMTL, though more candidate GPs used, the RMSE of estimated locations even become larger and show no convergence, which is because the both of them have no ability to eliminate the dictionary mismatch caused by the unknown PLE.  
It is also shown that although the performance of M-GEMTL is inferior to the proposed method, it is still better than M-SBL, which can be explained by its capability to infer the proper candidate GPs $\mathcal{G}$, and thus to a certain degree it can alleviate the dictionary mismatch caused by the mismatched initial candidate GPs .

This simulation suggests that the common thought that finer candidate grid leads to higher localization accuracy may not hold for all the sparsity-based MSL methods especially in the presence of uncertain propagation parameters, unknown nonuniform noise, and off-grid sources. It is also demonstrated that owing to the inference of the unknown propagation parameters, the proposed method can effectively take advantage of the finer grid granularity and thus show better performance.

%
%As a result, we can get from this simulation that the MSL problem is rather a joint parametric sparsifying dictionary learning and sparse signal recovery problem than a simple sparse recovery problem. Those methods which can partly or fully infer the dictionary parameters, e.g. M-GEMTL or the proposed PSBDL, can achieve better localization performance. And those methods able to learn all the dictionary parameters properly, such as the proposed PSBDL, can take advantage of the finer grid granularity to the best. Meanwhile, we are also suggested by the proposed method that it is possible and feasible to  learn the propagation parameters from the RSS observations, rather than the traditional routine to fit it from the enormous time-and-labor consuming off-line calibration  efforts.

\subsection{Impacts of  Measurement Perturbation}

\begin{table}[t]
	\newcommand{\tabincell}[2]{\begin{tabular}{@{}#1@{}}#2\end{tabular}}
	\centering
	%\setlength{\abovecaptionskip}{0pt}%
	%\setlength{\belowcaptionskip}{10pt}%
	%	\caption{\label{tab:11}The performance of four graph learning methods from signals generated from a random geometric graph.}
	\caption{RMSE of PLE estimate versus SNR (unknown nonuniform case)}
	\vspace{-0.2cm}
	%\Xhline{1.2pt}
	%\setlength{\abovecaptionskip}{0pt}
	%\setlength{\belowcaptionskip}{10pt}
	\begin{tabular}{ c c c c c c c }
		%\Xhline{1.2pt}
		\toprule
		%\hline
		%\specialrule{0em}{2pt}{2pt}
		%\tabincell{c}
		\textbf{SNR}[dB]   &  \tabincell{c}{0} & \tabincell{c}{2 } & \tabincell{c}{4 } &\tabincell{c}{6}  & \tabincell{c}{8} & \tabincell{c}{10}\\
		\midrule
		\textbf{CRLB}    &0.2416    &0.1869    &0.1465    &0.1217    &0.0885    &0.0380  \\
		%40 &0.7154   &0.4678   &0.8280   &2.0313  & 9.2731    \\
		\textbf{PSBDL}    &0.5711    &0.3983    &0.2568    &0.1924    &0.1258    &0.0926   \\
		\bottomrule	
	\end{tabular}
	\label{tab:PLE_SNR}
\end{table}  

\begin{figure}[t]
	\setlength{\abovecaptionskip}{-0.05cm}
	\begin{minipage}[b]{0.49\linewidth}
	    \centerline{\includegraphics[height = 7.5cm]{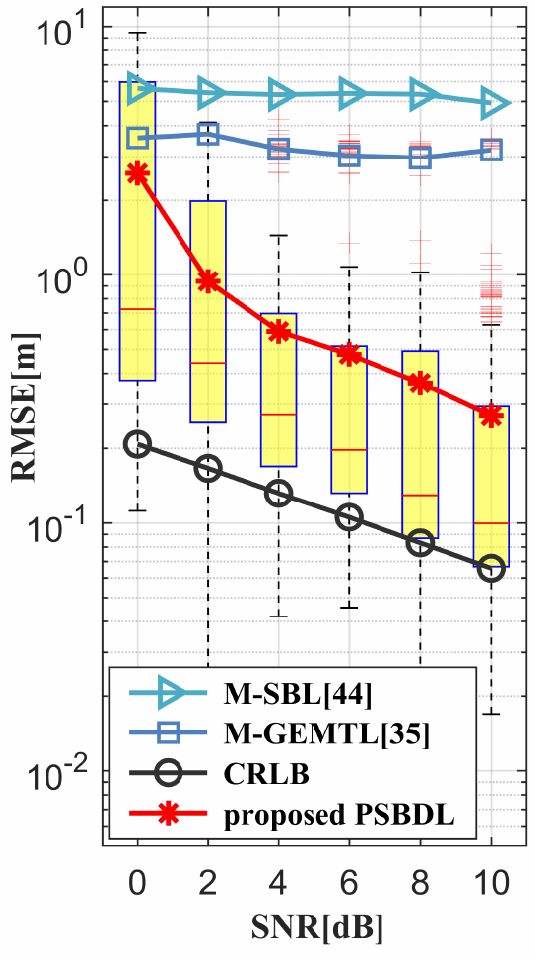}}
	    \vspace{-0.1cm}
		\centerline{(a)}
	\end{minipage}
	\begin{minipage}[b]{0.49\linewidth}
	   \centerline{\includegraphics[height = 7.5cm]{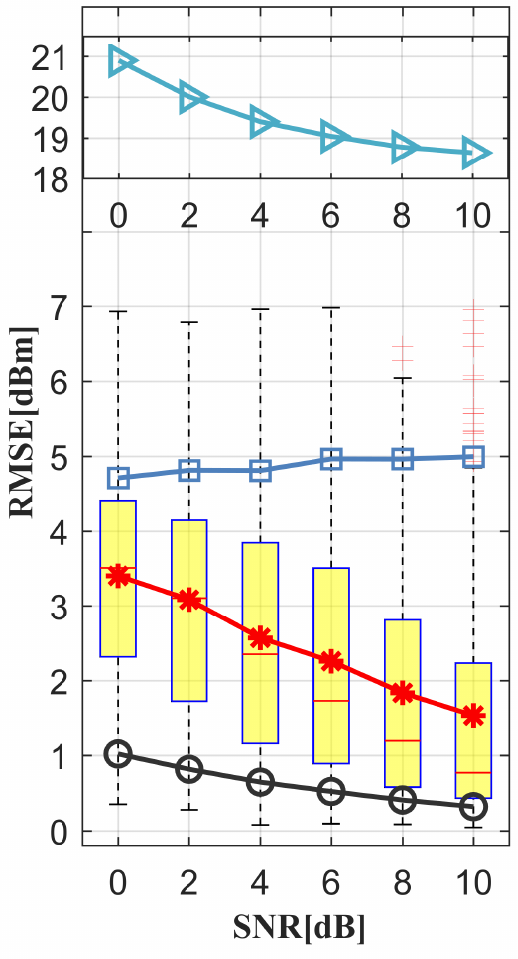}}
		\vspace{-0.1cm}
		\centerline{(b)}
		%\medskip
	\end{minipage}
	\caption{The RMSE of different  approaches versus SNR in the unknown nonuniform noise case for (a) the locations estimate and (b) the transmitted powers estimate.  }
	\label{fig:SNR}
	%	\vspace{-0.4cm}
\end{figure}

Secondly, as it is important to investigate the MSL methods under different measurement perturbation level, in this simulation, we set 
$N=121$, $M = 60$, $T=5$, and  consider the unknown nonuniform noise case.
%then  the  shadow fading case where the RSS measurement $y_{i}(t)$ is considered as log-normal random variable as  $ 10\lg\left( y_{i}(t) \right)  \sim \mathcal{N} \left( 10\lg\left(  \sum \nolimits_{ k = 1}^K {P_k} {f(\bm{s}_i,\bm{t}_k,\gamma_{ik} )} \right) ,\xi^2 \right) $ with the unit of the standard deviation $\xi$ being dB. 
Besides, since the mean of RMSE is susceptible to outliers, and may exaggerate the estimation error, box-plot is further provided henceforth for the proposed method to display the dispersion degree, the skewness, and the outliers  of its estimation errors. 

Tab.\ref{tab:PLE_SNR} shows the  RMSE of PLE estimate for  the proposed method  when SNR varies from 0dB to 10dB. As can be observed, the RMSE for PLE estimate of the proposed method decreases and gradually approach the theoretical CRLB when SNR increases, which verifies the effectiveness of the proposed method to retrieve PLE information from the observations under different noise levels.

Fig.\ref{fig:SNR}  presents the RMSE of the locations and the transmitted powers estimations for different approaches.   
It is clear in  Fig.\ref{fig:SNR} that the proposed method is superior to other approaches and its RMSE is close to the CRLB, which can be attributed to the joint inference of PLE and the proper candidate GPs $\mathcal{G}$. 
Specifically, the RMSE of M-SBL and M-GEMTL slightly decreases as SNR increases from 0dB to 10dB, whereas for the proposed method, its RMSE significantly decreases and show the same trend as the CRLB. Moreover, the box-plot discloses more information about the estimation errors. In  Fig.\ref{fig:SNR}(a), the second quartile or the median (red bar inside the box) is much more close to the first quartile (lower bound of the box) than the third quartile (upper bound), and the average RMSE is near the third quartile, which means the RMSE for location estimate is skewed-left and  is below the average RMSE  in about 75\% trails. Similarly, we can see from  Fig.\ref{fig:SNR}(b) that the RMSE of the proposed method for transmitted powers estimate is also skewed-left and more than half the trails have estimation error lower than the average RMSE.

This simulation underlines the effectiveness of the proposed method to joint learn the localization dictionary parameters and the sparse representation under different noise levels, which  can greatly improve the localization performance  and the robustness against the measurement perturbation.

%Fig.\ref{fig:SNR_shadow} plots the results of different levels of unknown nonuniform noise and shadow fading. It is observed that the shadow fading terms can be absorbed into measurement perturbation terms. When the standard deviation of the shadowing is small, the performance is dominated by the measurement perturbation, which
%makes sense, as pointed out by [10], because for suf

%\begin{figure}[b]	
%	\vspace{0.1cm}
%	%    	\setlength{\abovecaptionskip}{-0.05cm}
%	\centering
%	\epsfig{figure=fig_shadow_d_2,width=6cm}
%	\caption{The RMSE of different approaches versus SNR and shadow fading level for RSS measurements modeled as obeying log-normal distribution in dBm; (a) the locations estimate and (b) the transmitted powers estimate. }
%	\label{fig:SNR_shadow}
%	%	\vspace{1cm}
%\end{figure}

\begin{figure}
	\setlength{\abovecaptionskip}{-0.05cm}
	\begin{minipage}[b]{0.49\linewidth}
	    \centerline{\includegraphics[height = 7.5cm]{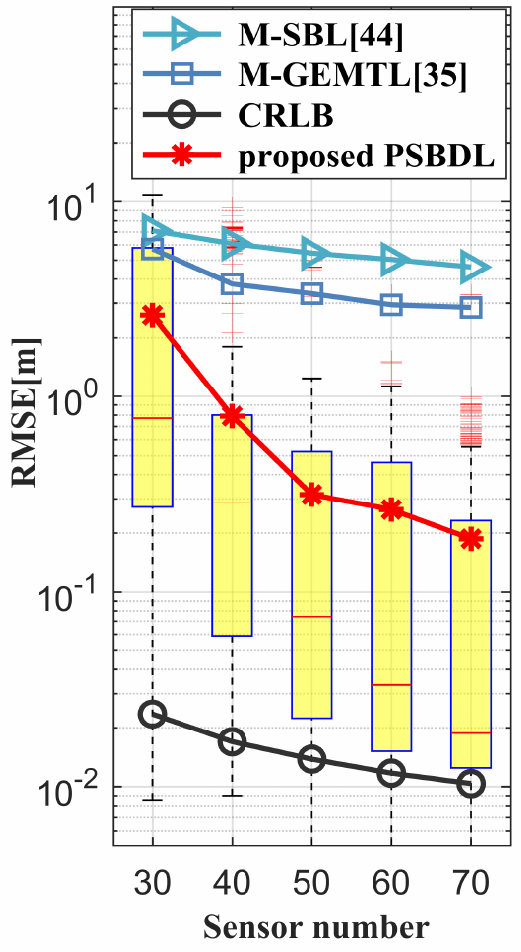}}
		\vspace{-0.1cm}
		\centerline{(a)}
	\end{minipage}
	\begin{minipage}[b]{0.49\linewidth}	
	    \centerline{\includegraphics[height = 7.5cm]{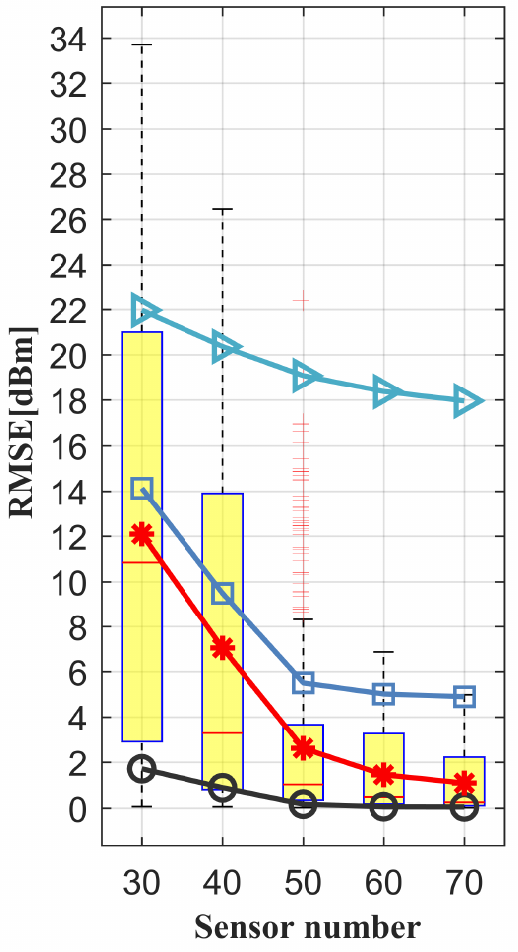}}
		\vspace{-0.1cm}
		\centerline{(b)}
		%\medskip
	\end{minipage}
	\caption{The RMSE of different approaches versus the number of sensors; (a) the locations estimate and (b) the transmitted powers estimate.  }
	\label{fig:sensor}
	%	\vspace{-0.4cm}
\end{figure}

\subsection{Effects of the Number of Sensors and Time Snapshots}
Finally, a natural method to improve the localization accuracy is to obtain more observations, i.e., to deploy more sensors and to gather more snapshot measurements.
In this simulation, we study the effects of  different numbers of sensors and time snapshots . 

Fig.\ref{fig:sensor} presents the RMSE of the locations and transmitted powers estimates for different
approaches when SNR = 25dB, $N$ = 121, $T$ = 5, and sensor number $M$ varying from 30 to 70.
It is observed that as the number of sensors increases, the RMSE of all algorithms decreases,
which is reasonable since more information obtained, less estimate uncertainty can be achieved. 
Furthermore, compared with other methods, the proposed PSBDL exhibits appreciably better accuracy with the same sensor number, requires fewer sensors under the same RMSE level, and approaches the CRLB.
More importantly, by comparing M-SBL, M-GEMTL, and the proposed PSBDL, we can conclude that although more sensors lead to less RMSE whether it infers the localization dictionary parameters or not, the more unknown dictionary parameters, e.g. the path-loss exponent, are effectively inferred, the more accurate estimation we obtain for the same number of sensors. 

\begin{figure}[t]
	\setlength{\abovecaptionskip}{-0.05cm}
	\begin{minipage}[b]{0.49\linewidth}
		  \centerline{\includegraphics[height = 7.5cm]{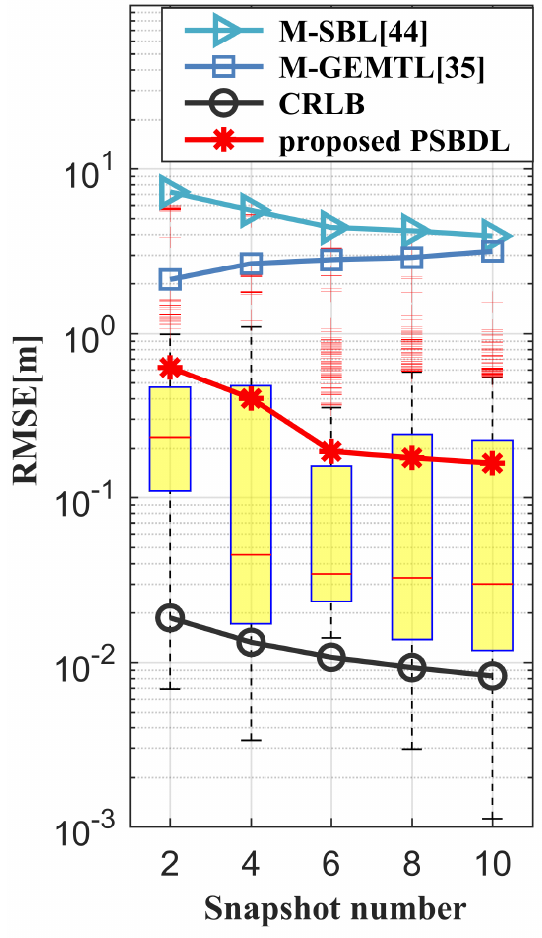}}
			\vspace{-0.1cm}
		\centerline{(a)}
	\end{minipage}
	\begin{minipage}[b]{0.49\linewidth}
	 \centerline{\includegraphics[height = 7.5cm]{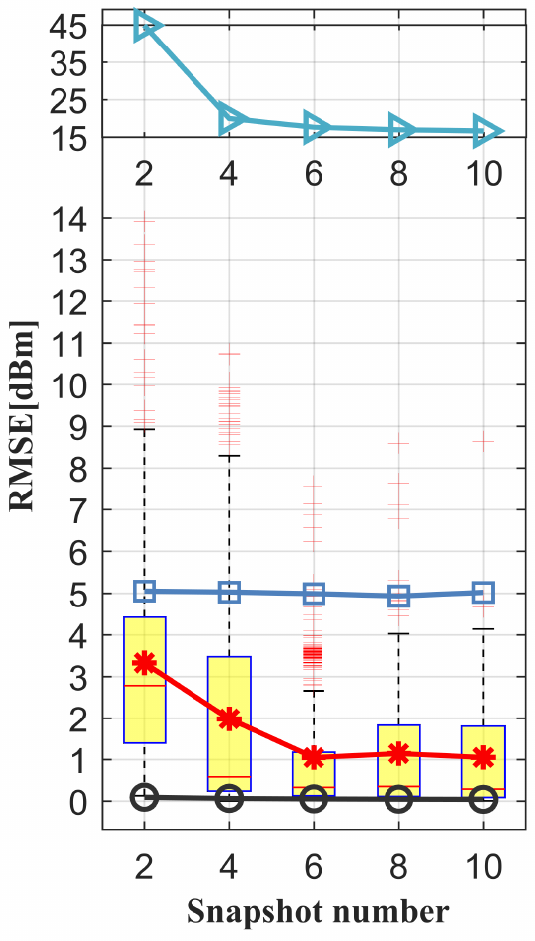}}
		\vspace{-0.1cm}
		\centerline{(b)}
		%\medskip
	\end{minipage}
	\caption{The RMSE of different approaches versus the number of snapshots; (a) the locations estimate and (b) the transmitted powers estimate.  }
	\label{fig:snapshot}
	%	\vspace{-0.4cm}
\end{figure}

Fig.\ref{fig:snapshot} plots the RMSE results of the locations and transmitted powers estimates for different approaches with SNR = 25dB, $N$ = 121, $M$ = 60, and snapshot number $T$ changing from 2 to 10. It is shown that the proposed method can effectively exploit the gains of more snapshots, and exhibits the best performance under all different snapshot numbers, which indicates the importance of the inference of the localization dictionary parameters and the sparse representation jointly.

Interestingly, different from Fig.\ref{fig:sensor},  the RMSE curve of the  proposed method rapidly converges with the number of snapshots increases, and more snapshots do not significantly improve the performance of M-GEMTL, which suggests that 1) more snapshots will improve the localization accuracy but with limited ability compared with more sensors, and 2) the algorithm should be carefully designed to effectively utilize the information gains of more snapshot data. 

This simulation indicates that the proposed method can effectively exploit the gains of  more sensors and more snapshots which will contribute to the improvement of estimation accuracy.

\section{Discussion}
In this section, we first provide a Bayesian gain interpretation to further understand the proposed algorithm, and then clarify its difference with some well-known dictionary learning algorithms.

\vspace{-0.2cm}
\subsection{Bayesian Gain Interpretation for the Proposed Algorithm} 
\begin{figure}[b]	
	\vspace{0.1cm}
	  \centerline{\includegraphics[width=8cm,height = 6cm]{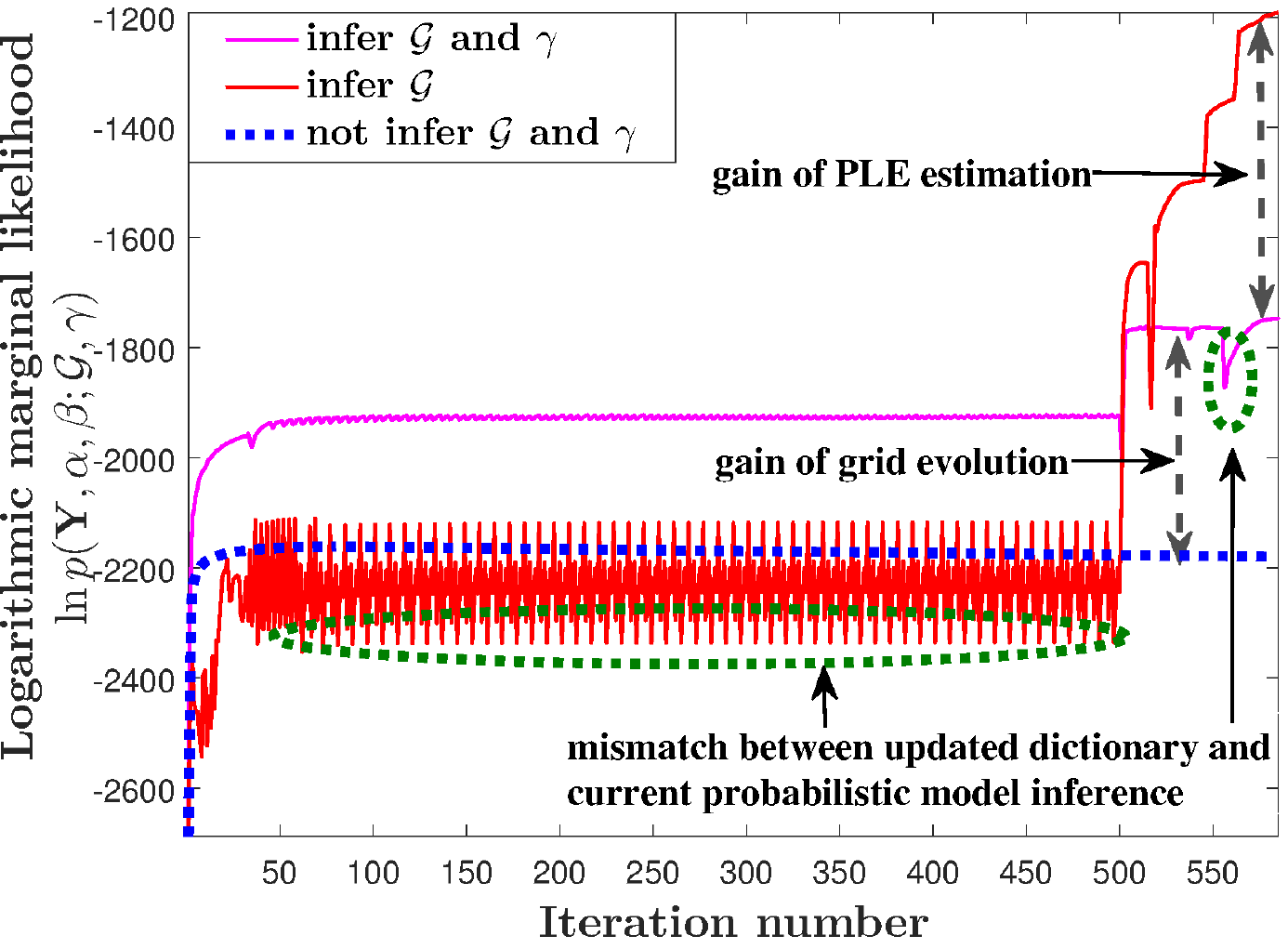}}
	\caption{Changes of the logarithmic marginal likelihood  along with iterations.}
	\label{fig:MLProcess}
	%	\vspace{1cm}
\end{figure}

To further understand the mechanism of the proposed algorithm, we illustrate  by Fig. \ref{fig:MLProcess} the Bayesian gains under different treatments  from the perspective of Bayesian inference.     
The logarithmic marginal likelihood   $\ln p({\bm{Y}},\bm{\alpha},\bm{\beta}; \mathcal{G},\gamma)$  serves as not only the goal function, but also the measurement of matching degree among the observation, the probabilistic model inference, and  the localization dictionary. 

Note that for all treatments  in  Fig. \ref{fig:MLProcess}, both the true grid $\bar{\mathcal{G}}$ and the true PLE $\bar\gamma$ is unknown, and the initial values $\mathcal{G}^{(0)}$ and $\gamma^{(0)}$  differ from the truth.
It is obvious that inferring $\mathcal{G}$ and $\gamma$ jointly (the proposed treatment) brings much higher likelihood than only inferring $\mathcal{G}$ (state-of-the-art off-grid MSL treatment, e.g. \cite{GEMTL,SUN_CL_2017} ), which can be attributed to the  gain of PLE estimation, while the likelihood of the latter is higher than not inferring the localization dictionary parameter $\mathcal{G}$ and $\gamma$ (conventional CS-based treatment, i.e. \cite{Zhang_2011, Feng_2012,MSBL}), which is owing to the gain of grid evolution.    
More specifically,  if we do not infer the localization dictionary (namely $\mathcal{G}$ and $\gamma$), the likelihood curve, shown by the blue dotted line, exhibits the typical smooth convergence curve of EM algorithm, while the curves of all other treatment, shown by magenta and red solid lines, grow in stages. Those stages are internal iterations of the algorithms, and the likelihood may drop at the edges of those stages, owing to the mismatch between the updated dictionary model and current probabilistic inference.

\vspace{-0.2cm}
\subsection{Difference with Some Dictionary Learning Problems}
Problem (\ref{JointOpt}) may have other interpretations such as underdetermined blind source separation \cite{SCA} and  other CS dictionary learning problems, such as K-SVD \cite{KSVD}. The difference is that in these works, the coefficient matrix or dictionary matrix $\bm{\Phi}$ need not have a certain physical structure and the only requirement is that $\bm{Y}$ has sparse representation under $\bm{\Phi}$, while for the localization problem here, $\bm{\Phi}$ are physical structured, or more precisely, parameterized by the dictionary parameters of interest according to a certain physical model. 

Mathematically, the original MSL problem shown in Section \ref{Section:MSL_model} is a  highly nonlinear and nonconvex  continuous multi-parameter  optimization problem, which is hard to solve directly. In this paper, like piecewise linear approximation, we use a series of discrete parametric dictionary model to iteratively approximate the original complex continuous optimization problem, and reformulate the original problem as a sparse recovery problem under the parametric discrete dictionary. In each iteration, we jointly infer the optimal step of current dictionary parameters to the truth and the sparse representation under current dictionary, which is implemented by the incorporation of dictionary approximation model and sparse Bayesian learning framework. Hence, we termed the proposed method as  parametric sparse Bayesian dictionary learning.

\section{Conclusion}

In this paper, we have investigated the multiple co-channel sources localization problem based on RSS measurements in the presence of unknown nonuniform measurement perturbations and  uncertain propagation parameters including both the transmitted powers and the path-loss exponent. The original MSL problem is highly nonconvex and hard to solve. With the combination of the sparsity-based MSL
model and the  localization dictionary approximation model, we have reformulated the original MSL problem into a joint PSDL and SSR problem which was solved by the proposed PSDBL method.
Extensive simulations were carried out compared with the state-of-the-art sparsity-based MSL methods and the theoretical CRLB we derived. Numerical results highlighted the 
effectiveness and superiority of the proposed method, and also shed light on its importance and  feasibility of jointly inferring from the  RSS measurements the source locations and the propagation parameters. Mathematically, this paper provides a paradigm to enforce sparse representation and  approximate a continuous sparsifying parametric dictionary by  a series of discrete parametric dictionary simultaneously.

% or


\appendices

\section{Derivation of (\ref{update_alpha})   } \label{AppendixC}
To obtain Eq. (\ref{update_alpha}), first let $ Q(\bm\alpha) = E \left\{\ln p({\bm{X}}|\bm{\alpha})p(\bm{\alpha} )   \right\} $. Then we have
\begin{align}
	& Q(\bm{\alpha})  \notag \\
	&\!=\! E\left\{  - \frac{1}{2}\sum\limits_{t = 1}^T {\left( {\ln \left| {\bm{A}} \right| + \bm{x}{{(t)}^T}{{\bm{A}}^{ - 1}}\bm{x}(t)} \right)}  + \sum\limits_{i = 1}^N {\left( {\ln \lambda  - \frac{\lambda }{2}{\alpha _i}} \right)}  \right\} \notag  \\
	& \quad\quad + const \notag  \\
	& \!=\! - \frac{1}{2}\sum\limits_{t = 1}^T {\left\{ \sum\limits_{i = 1}^N {\ln {\alpha _i}}  \!+\! \sum\limits_{i = 1}^N  {\alpha _i^{ - 1} \left( {\mu}(t)_i^2 + \Sigma_{ii} \right) } \right\} }   \notag\\
	&\quad+ \sum\limits_{i = 1}^N {\left( {\ln \lambda  - \frac{\lambda }{2}{\alpha _i}} \right)}  + const  
\end{align}
with $const$ being the item constant to $\bm{\alpha}$. To find the stationary point of $ Q(\bm\alpha)$ w.r.t $\alpha_i$, let $\partial{ Q(\bm\alpha)}/\partial{\alpha_i} = 0 $, and then we obtain
\begin{equation}
	\alpha _i 	=\!\frac{\sqrt{ T^2 + 4\lambda \sum\nolimits_{t=1}^{T}{\left( {\Sigma}_{ii} + {\mu}(t)_i^2 \right)}  } -T}{2\lambda}.
\end{equation}

\section{Derivation of (\ref{update_beta})   } \label{AppendixD}
Let  $ Q(\bm\beta) =  E \left\{ p({\bm{Y}}|{\bm{X}},\bm{\beta},\gamma;\mathcal{G}) p(\bm{\beta})   \right\} $. 
Then, we have
\begin{align}
	& Q(\bm\beta) \notag \\
	&  = E\Bigg\{ \frac{1}{2}\sum\limits_{t = 1}^T {\left( {\ln \left| \bm{B} \right| - {{\left( {\bm{y}(t) - \bm{\Phi} \bm{x}(t)} \right)}^T} \bm{B}\left( {\bm{y}(t) - \bm{\Phi} \bm{x}(t)} \right)} \right)} \notag\\
	&\quad + \sum\limits_{j = 1}^M {\left( {\left( {a - 1} \right)\ln {\beta _j} - b{\beta _j}} \right)}  \Bigg\} + const \notag\\
	&  = \frac{1}{2}\sum\limits_{t = 1}^T {\left( {\ln \left| \bm{B} \right| - \sum\limits_{j = 1}^M {{\beta _j}E\left\{ {\left( {\bm{y}(t) - \bm{\Phi} \bm{x}(t)} \right)_j^2} \right\}} } \right)}   \notag\\
	& \quad + \left( {a - 1} \right)\ln \left| \bm{B} \right| - b\sum\limits_{j = 1}^M {{\beta _j}} + const \notag\\
	&  = \left( {a - 1 + \frac{T}{2}} \right)\sum\limits_{j = 1}^M {\ln {\beta _j}}  \notag\\
	&  \quad   - \frac{1}{2} \sum\limits_{j = 1}^M {{\beta _j}\sum\limits_{t = 1}^T {E\left\{ {\left( {\bm{y}(t) - \bm{\Phi} \bm{x}(t)} \right)_j^2} \right\}} } - b\sum\limits_{j = 1}^M {{\beta _j}} +const
\end{align}
where $const$ is the item constant to $\bm{\beta}$.	let $\partial{ Q(\bm\beta)}/\partial{\beta_j} = 0 $, we have
\begin{equation} \label{update_beta_temp}
	{\beta_j} = \frac{{2a - 2 + T}}{{2b + \sum\nolimits_{t=1}^{T}E\left\{ {\left( {\bm{y}(t) - \bm{\Phi} \bm{x}(t)} \right)_j^2} \right\}   }}
\end{equation}

Since $\bm{y}(t) - \bm{\Phi} \bm{x}(t)\sim \mathcal{N}\left( {\bm{y}(t) - \bm{\Phi} \bm{\mu} (t),\bm{\Phi} \bm{\Sigma} {\bm{\Phi} ^T}} \right)$, denote by $\bm{e}_j$ the unit column vector with its $j$-th element being one, then we have
\begin{align} \label{E_residual}
	&E\left\{ {\left( {\bm{y}(t) - \bm{\Phi} \bm{x}(t)} \right)_j^2} \right\} \notag\\
	&= E\left\{ {{{\left( {\bm{y}(t) - \bm{\Phi}\bm{x}(t)} \right)}^T}{\bm{e}_j}\bm{e}_j^T\left( {\bm{y}(t) - \bm{\Phi} \bm{x}(t)} \right)} \right\} \notag\\
	&=tr\left( {{\bm{e}_j}\bm{e}_j^T \bm{\Phi} \bm{\Sigma} {\bm{\Phi} ^T}} \right) + {\left( {\bm{y}(t) - \bm{\Phi} \bm{\mu} (t)} \right)^T}{\bm{e}_j}\bm{e}_j^T\left( {\bm{y}(t) - \bm{\Phi} \bm{\mu} (t)} \right)\notag\\
	& = tr\left( {{\bm{e}_j^T \bm{\Phi} \bm{\Sigma} {\bm{\Phi} ^T} \bm{e}_j}} \right) + {\left( {\bm{y}(t) - \bm{\Phi} \bm{\mu} (t)} \right)_j^2}\notag\\
	&=   \left(\bm{\Phi} \bm{\Sigma} \bm{\Phi}^T\right)_{jj} +  {\left( {\bm{y}(t) - \bm{\Phi} \bm{\mu} (t)} \right)_j^2}.
\end{align} 

Finally, substituting (\ref{E_residual}) into Eq.(\ref{update_beta_temp}), we obtain the update formula shown by  Eq.(\ref{update_beta}).

\section{Derivation of (\ref{E_step_theta})   } \label{AppendixA}
Eq.(\ref{E_step_theta}) is based on the following properties:
if $\bm{x} \sim \mathcal{N} \left(\bm{\mu}, \bm{\Sigma} \right)$, then we have
\begin{equation}
	E \left\{  \bm{x}^{T}\bm{A} \bm{x}  \right\} =  \bm{\mu}^{T}\bm{A} \bm{\mu}  + \operatorname{tr}\left(\bm{A}\bm{\Sigma}\right).
\end{equation}
With this property,  we obtain 	Eq.(\ref{E_step_theta}) as
\begin{align}
	&E\left\{ \sum_{t=1}^{T} \left( \bm{y}(t) \!-\! \bm{\Phi} \bm{x}(t)\right)^{T} \bm{B} \left( \bm{y}(t) \!-\! \bm{\Phi} \bm{x}(t)\right) \right\} 
	\notag\\
	&=\!\! \sum\limits_{t = 1}^T \!\!\Bigg\{
	\bm{y}(t)^{T}\bm{B}\bm{y}(t) \!-\! 2\bm{\mu}(t)^{T}\bm{\Phi}^{T}\bm{B}\bm{y}(t) \!+\! E\left\{\bm{x}(t)^{T}\bm{\Phi}^{T}\bm{B}\bm{\Phi}\bm{x}(t) \right\} \!\Bigg\}
	\notag\\
	&=\!\! \sum\limits_{t = 1}^T \!\!\Bigg\{	\left( \bm{y}(t) \!-\! \bm{\Phi} \bm{\mu}(t)\right)^{T} \bm{B} \left( \bm{y}(t) \!-\! \bm{\Phi} \bm{\mu}(t)\right) \!+\! \operatorname{tr}\left(\bm{\Phi} \bm{\Sigma}\bm{\Phi} ^{T} \bm{B}\right) \!\Bigg\}.
\end{align}

\section{Derivation of (\ref{delta_LLSQ}) } \label{AppendixB}

To obtain the goal function in (\ref{delta_LLSQ:goal}), we first introduce the following matrix identities:
\begin{gather}
	\bm{v}^T \operatorname{diag}\left( \bm{u} \right) = \bm{u}^T \operatorname{diag}\left( \bm{v} \right) ,\label{MtxIdty1}\\
	\operatorname{diag}\left( \bm{v} \right) \cdot \bm{M} \cdot \operatorname{diag}\left( \bm{u} \right) = \bm{M} \circ \bm{v}\bm{u}^T,\label{MtxIdty2}\\
	\operatorname{tr} \left\{  \operatorname{diag}\left( \bm{v} \right)^H  \bm{Q} \operatorname{diag}\left( \bm{u} \right) \bm{R}^T  \right\}  = \bm{v}^H \left( \bm{Q} \circ \bm{R} \right) \bm{u},  \label{MtxIdty3}
\end{gather}
for  vector $\bm{v}$, $\bm{u}$,  and  matrices $\bm{Q}$ and $\bm{R}$ with proper dimension, where $(\cdot)^H$ denotes the  conjugate transpose operator. 

Then, substituting (\ref{iter_subq:c2}) into Eq. (\ref{E_step_theta}),  by (\ref{MtxIdty1}) and (\ref{MtxIdty2})  we have the first term inside the RHS of  Eq. (\ref{E_step_theta}) expressed as
\begin{small}
	\begin{align}
		&\!\!\!\!\!\left( \bm{y}(t) \!-\! \bm{\Phi} \bm{\mu}(t)\right)^{T} \bm{B} \left( \bm{y}(t) \!-\! \bm{\Phi} \bm{\mu}(t)\right)   \notag\\
		&\!\!\!\!\! = \delta_{\gamma}^2 \left(  {\bm{\Phi}'_{\gamma}} \bm{\mu}(t) \right)^{T}\bm{B} {\bm{\Phi}'_{\gamma}} \bm{\mu}(t)  + \!\!\sum\limits_{\chi = u,v } \!\!{\bm{\delta}_\chi^T} \left( {\bm{\Phi}'_{\chi}}^T \bm{B}{\bm{\Phi}'_{\chi}} \circ {\bm{\mu}(t)} {\bm{\mu}(t)} ^T\right) {\bm{\delta}_\chi}  \notag\\
		& \!\!\!\!\! +\! 2{\bm{\delta}_u^T} \!\! \left( {\bm{\Phi}'_{u}}^T \bm{B} {\bm{\Phi}'_{v}} \!\circ\!  {\bm{\mu}(t)} {\bm{\mu}(t)} ^T\right)\! {\bm{\delta}_v} \!+\! 2 \delta_{\gamma} \!\!\!\!\sum\limits_{\chi = u,v } \!\!\!\! {\bm{\delta}_\chi^T} \!\!\left( {\bm{\Phi}'_{\chi}}^T \bm{B}  {\bm{\Phi}'_{\gamma}} \! \circ \! {\bm{\mu}(t)} {\bm{\mu}(t)} ^T\right) \! {\bm{1}_N}\notag\\
		&\!\!\!\!\! -\! 2 {{\delta}_\gamma} {\left( { \bm{y}(t) - {\bm\Phi}_0 \bm{u}(t)} \right)^T} \bm{B} {\bm{\Phi}'_{\gamma}}  \operatorname{diag}(\bm{u}(t)) \notag\\
		& \!\!\!\!\!- 2 \sum\limits_{\chi = u,v} {\left( { \bm{y}(t) - {\bm\Phi}_0 \bm{\mu}(t)} \right)^T} \bm{B}  {\bm{\Phi}'_{\chi}}  \operatorname{diag}(\bm{\mu}(t)) {\bm{\delta}_\chi} + const,	
	\end{align}
\end{small}

\noindent and by  (\ref{MtxIdty3}), we obtain the second term inside the RHS of  Eq. (\ref{E_step_theta}) as
\begin{small}
	\begin{align}
		&\!\!\!\!\!\!\!\operatorname{tr}\left\{ \left({\bm\Phi}_0 \!+\!  {\bm{\Phi}'_{u}}\bm{\Delta}_{u}  \!+\!  {\bm{\Phi}'_{v}}\bm{\Delta}_{v}  \!+\!  \delta_{\gamma}{\bm{\Phi}'_{\gamma}}\right)  \bm{\Sigma} \left({\bm\Phi}_0 \!+\!  {\bm{\Phi}'_{u}}\bm{\Delta}_{u}  \!+\!  {\bm{\Phi}'_{v}}\bm{\Delta}_{v} \!+\!  \delta_{\gamma}{\bm{\Phi}'_{\gamma}}\right) ^{T} \bm{B}\right\} \notag\\
		&\!\!\!\!\!\!\!= \delta_{\gamma}^2 \; tr \!\left\{ {\bm{\Phi}'_{\gamma}}  \Sigma {\bm{\Phi}'_{\gamma}}^T \bm{B} \right\}  \! + \!\!\!\! \sum\limits_{\chi = u,v} \!\!\!\! \bm{\delta}_{\chi}^T \!\left( {\bm{\Phi}'_{\chi}}^T  \bm{B} \bm{\Phi}'_{\chi} \!\circ\!  \bm{\Sigma} \right) \! \bm{\delta}_{\chi}  \!+\!  2{\bm{\delta}_u^T} \!\! \left( {\bm{\Phi}'_{u}}^T \bm{B} {\bm{\Phi}'_{v}} \!\circ\! \bm{\Sigma} \right)\! {\bm{\delta}_v} \notag\\
		&\!\!\!\!\!\!\! + 2 \delta_{\gamma} \!\! \sum\limits_{\chi = u,v } \!\!{\bm{\delta}_\chi^T} \!\!\left( {\bm{\Phi}'_{\chi}}^T \bm{B} {\bm{\Phi}'_{\gamma}} \! \circ \! \bm{\Sigma} \right) \! {\bm{1}_N}	 + 2 \!\!\! \sum\limits_{\chi = u,v} \!\! \operatorname{diag}\left(  {\bm{\Phi}'_{\chi}} \bm{B}\bm{\Phi}_0 \bm{\Sigma} \right)^T \bm{\delta}_{\chi} \notag \\
		&\!\!\!\!\!\!\! + 2 \delta_{\gamma} \operatorname{tr}\left\{ \bm{\Phi}_0 \bm{\Sigma} {\bm{\Phi}'_{\gamma}}^T \bm{B}\right\} + const
	\end{align} 
\end{small}

\noindent where $const$ is constant independent of $\bm{\delta}_{u}$, $\bm{\delta}_{v} $ and $\delta_{\gamma}$. Finally, by plugging these terms into (\ref{E_step_theta}), we obtain the goal function in (\ref{delta_LLSQ}).

%	\section*{Acknowledgment}
%	The authors would like to thank the anonymous reviewers and the associate editor for their useful comments that significantly improved the quality of this paper.

\ifCLASSOPTIONcaptionsoff
\newpage
\fi



%\bibliographystyle{IEEEtran}
%
%% argument is your BibTeX string definitions and bibliography database(s)
%\bibliography{IEEEabrv,xxx.bib}
\bibliographystyle{IEEEtran} 
\begin{footnotesize}
	\bibliography{IEEEabrv,my_reference} 
\end{footnotesize}

\end{document}